\documentclass[aps,preprintnumbers,eqsecnum,amsmath,amssymb,nofootinbib]{revtex4}
\usepackage{graphicx}
\usepackage[english]{babel}
\usepackage{url}
\usepackage{graphicx}
\usepackage{xcolor}
\usepackage{amsmath}
\usepackage{amssymb}
\usepackage{slashed}
\usepackage{multirow}
\begin{document}
\title{Probing vector- vs. scalar-mediator dark-matter scenarios in $B\to (K,K^*) M_X$ decays}
\author{Alexander Berezhnoy$^{a}$, Wolfgang Lucha$^{b}$ and Dmitri Melikhov$^{a,c,d}$}
\affiliation{
$^a$D.~V.~Skobeltsyn Institute of Nuclear Physics, M.~V.~Lomonosov Moscow State University, 119991, Moscow, Russia\\
$^b$Institute of High Energy Physics, Austrian Academy of Sciences, Nikolsdorfergasse 18, A-1050 Vienna, Austria\\
$^c$Joint Institute for Nuclear Research, 141980 Dubna, Russia\\
$^d$Faculty of Physics, University of Vienna, Boltzmanngasse 5, A-1090 Vienna, Austria}
\date{\today}
\begin{abstract}
  Within the hypothesis of the dark-matter origin of the excess in $B\to K M_X$ decays over the standard-model expectation, 
  observed by Belle-II, we show that:
  (i) Scalar- and vector-mediator scenarios may be unambiguously discriminated by measuring
  the differential distributions in $B\to K M_X$ and  $B\to K^* M_X$ decays. (ii) Combining the available data on
  $\Gamma(B\to K M_X)$ and the upper limit on $\Gamma(B\to K^* M_X)$ provides a tight constraint on the vector-mediator mass $M_V\lesssim 3$ GeV. At the same time, no constraints on the scalar-mediator mass are imposed by these data.
  (iii) Both scalar- and vector-mediator scenarios allow a good description of the differential distributions in $B\to K M_X$
  measured by Belle-II and an extraction of dark-model parameters within both scenarios. 
\end{abstract}
\maketitle

\section{\label{Sect:1}Introduction}
In the Standard Model (SM) $B^+\to K^+ M_X$ is reduced to $B^+\to K^+\bar\nu\nu$. 
This is a flavour-changing neutral-current (FCNC) induced process which is forbidden in the Standard Model (SM)
at the tree level and occurs via loops (penguins and boxes). It has been calculated with rather good accuracy
(see the recent papers \cite{damir2023epjc,hpqcd2023,Allwicher:2023xba}), yielding 
\begin{eqnarray}
  \label{damir}
Br(B^+\to K^+\bar\nu\nu)&=&(4.44\pm 0.30) \cdot 10^{-6},\nonumber\\
Br(B^+\to K^{*+}\bar\nu\nu)&=&(9.8\pm 1.4) \cdot 10^{-6}. 
\end{eqnarray}
(In these numbers, the contributions of the tree-level process $B\to \bar\nu (\tau\to \nu K)$ have been subtracted. 
This is the relevant quantity to be compared with
the experimental data, see \cite{Allwicher:2023xba}.)

In 2017 Belle reported the upper limits \cite{belle2017}
\begin{eqnarray}
Br(B\to K\bar\nu\nu)&<& 1.6\cdot 10^{-5},
\nonumber\\
Br(B\to K^*\bar\nu\nu)&<& 2.7\cdot 10^{-5},
\label{belle2017}
\end{eqnarray}
which agree with the SM. 
A great surprise came in 2023 when Belle-II reported the following observation \cite{Belle-II:2023esi} 
\begin{eqnarray}
Br(B^+\to K^+\bar\nu\nu)&=&(2.3\pm 0.7) \cdot 10^{-5}=(5.4\pm 1.5)Br(B^+\to K^+\bar\nu\nu)_{\rm SM},
\end{eqnarray}
well exceeding the SM expectation. Also the differential distribution in $B\to K M_X$ has been measured. 

Immediately after that, many attempts to explain the excess within different new-physics
scenarios have been
performed
\cite{Athron:2023hmz, Bause:2023mfe, Felkl:2023ayn, Abdughani:2023dlr,Dreiner:2023cms,He:2023bnk,datta,Berezhnoy:2023rxx,Calibbi:2025rpx,Altmannshofer:2025eor, Zhang:2024hkn,Bhattacharya:2024clv,Altmannshofer:2024kxb,Davoudiasl:2024cee,Kim:2024tsm,Marzocca:2024hua,Bolton:2024egx,He:2024iju,Hou:2024vyw,Gabrielli:2024wys,Loparco:2024olo,Ho:2024cwk,Fridell:2023ssf,McKeen:2023uzo,Altmannshofer:2023hkn,Wang:2023trd,He:2025jfc,fajfer,blm2025,Aliev:2025hyp,Kolay:2025jip,He:2025zfy,DiLuzio:2025whh,Ding:2025eqq}.

One of the discussed possibilities is the conjecture that the excess in $B\to K M_X$ is due to the decay into
a pair of invisible dark-matter (DM) fermions $\bar\chi\chi$ which couples to the SM particles via a mediator field $R$,
which may be a scalar ($R=\phi$) or a vector ($R=V$) field:
\begin{eqnarray}
\Gamma(B\to K M_X)=\Gamma(B\to K \bar\nu\nu)_{\rm SM}+\Gamma(B\xrightarrow{R}K \bar\chi\chi).
\end{eqnarray}
Within such scenarios, a similar excess is expected also in other decay modes $B\to (K^*,\pi,\rho)M_X$
compared to the SM predictions for $B\to (K^*,\pi,\rho)\bar\nu\nu$. 
In \cite{Berezhnoy:2023rxx}, we discussed a scalar-mediator scenario (S-scenario) and demonstrated that
within this scenario rigorous constraints on the differential rates of the excess events in $B\to K^* M_X$
may be obtained. In \cite{blm2025}, we showed that the S-scenario allows an excellent description of the differential
distributions of the excess events reported by \cite{Belle-II:2023esi}
and extracted the corresponding DM parameters. 

We now extend our analysis and include both the integrated and the differential
decay rates of the excess events in $B\to K M_X$ decays for S- and V-scenarios.
Among the main results of this paper we mention the following:

\noindent
(i) The ratios of the {\it differential} distributions
of the excess events in $B\to K^* M_X$ and $B\to K M_X$ decays have qualitatively different shapes in S- and V-scenarios,
independently of the parameters of the DM particles, thus providing an unambiguous probe of the mediator spin. 

\noindent
(ii) The ratios of the {\it integrated rates} of the excess events in $B\to K^* M_X$ and $B\to K M_X$ are
sensitive to the mediator mass and its width. Within the DM conjecture, combining the measured $\Gamma(B\to K M_X)$ and the
upper limit on $\Gamma(B\to K^* M_X)$ yields a tight constraint on the vector-mediator mass $M_V<3$ GeV. 
At the same time, no constraint on the scalar mediator mass $M_\phi$ is imposed by the existing data on the integrated rates. 

\noindent
(iii) Both S- and V-scenarios allow a good description of
the {\it differential distributions} in $B\to K\,M_X$ measured by Belle-II \cite{Belle-II:2023esi} and
an extraction of the corresponding DM parameters. 

The paper is organized as follows: In the next Section, we provide details of S and V dark-matter scenarios and
derive the corresponding expressions for the differential decay rates. Section \ref{Sect:3} presents important signatures
of S- and V-scenarios in differential and integrated rates of the excess events in $B\to K M_X$ and $B\to K^* M_X$ decays.
Section \ref{Sect:4} shows the results of the fits to the differential distributions of the excess events in $B\to K M_X$
as measured by Belle-II in S- and V-scenarios. Section \ref{Sect:5} gives our summary. 

\section{\label{Sect:2}Conjecture of dark-matter origin of the excess events in $B\to KM_X$}   
We are going to test the conjecture that the Belle-II excess events are due to the decay into dark-matter (DM)
fermions ($\chi$) which are coupled to the top quark via a mediator field $R$. No interaction of the
mediator with other SM particles is implied (such scenarios are referred to as {\it top-philic} mediator scenarios).
In this case, the FCNC $b\xrightarrow{R}s\bar\chi\chi$ decay proceeds via the top-$W$ loop depicted in Fig.~\ref{Fig:1}: 
\begin{center}
\begin{figure}[h]
  \includegraphics[width=6cm]{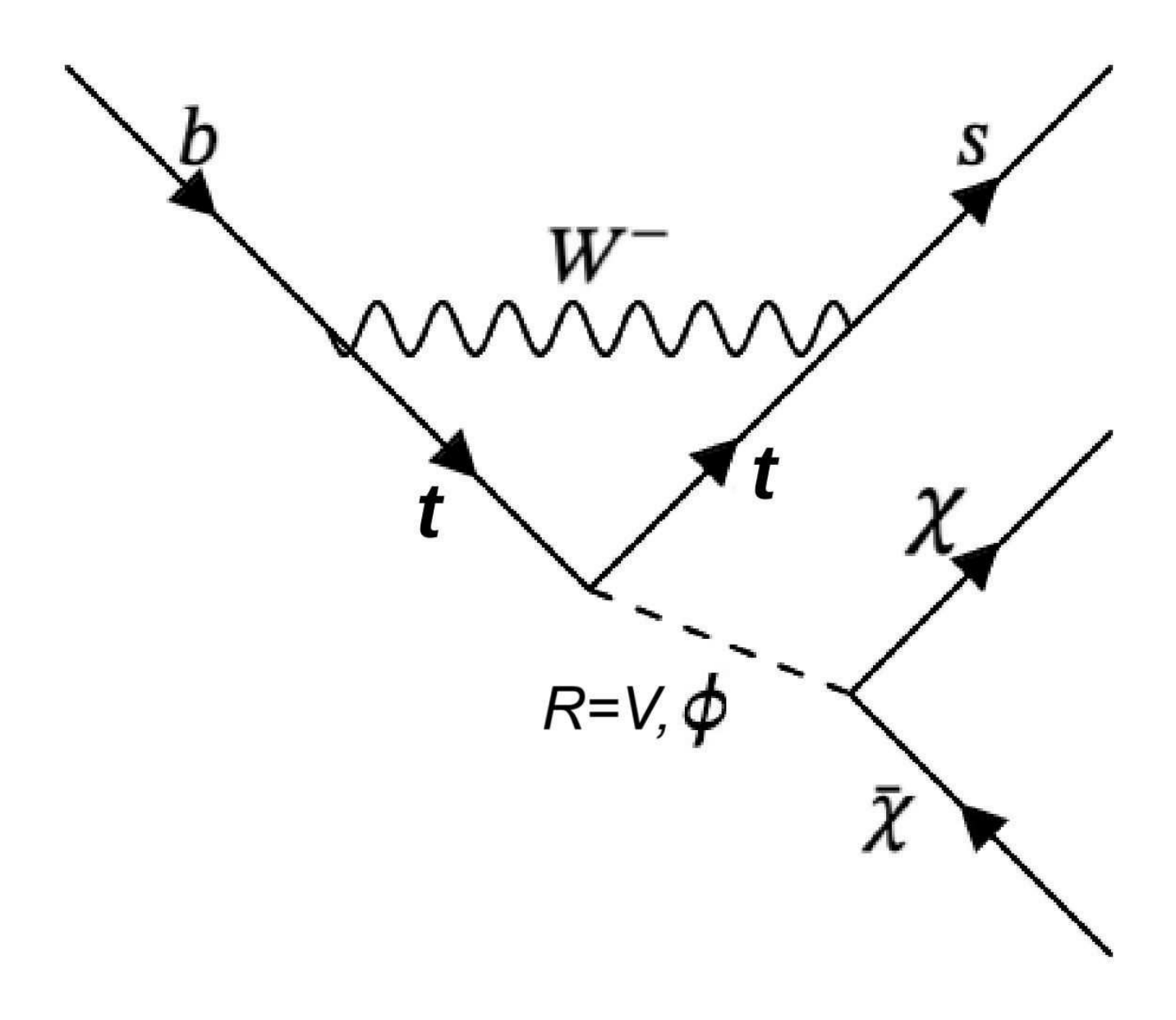}
  \caption{\label{Fig:1}The diagram describing the $b\to s\bar\chi\chi$ amplitude in a top-philic DM scenario}
\end{figure}
\end{center}
In this analysis, both a scalar mediator $R=\phi$ (S-scenario) and
a vector mediator $R=V$ (V-scenario) are considered. 

\subsection{S-scenario}
Let us briefly recall the analysis of \cite{Berezhnoy:2023rxx,blm2025}
of a rather popular model involving an
interaction of DM fermions $\chi$ with the top quark $t$ by exchange of a scalar-mediator field $\phi$,
governed by the interaction Lagrangian \cite{Batell:2009jf, Schmidt-Hoberg:2013hba}
\begin{eqnarray}
 \mathcal{L_{\rm int}} = - \frac{y m_t}{v} \phi\, \bar t t - g_{\phi\chi\chi}\phi \bar \chi \chi, 
  \label{eq:lagrangian}
\end{eqnarray}
where $v\simeq 246$ GeV is the Higgs vacuum expectation value and $y$
parametrizes the $\phi \bar tt$ coupling. 
The emerging effective Lagrangian encoding the FCNC vertex $b\to s\phi$ then reads
\cite{Batell:2009jf, Schmidt-Hoberg:2013hba}
\begin{eqnarray}
\label{Leff}
\mathcal{L}_{b \rightarrow s\phi} = g_{bs\phi}\,\phi\, \bar s_L b_R + {\rm h.c.}, \qquad
 g_{bs\phi}=\frac{y m_b}{v} \frac{3 \sqrt{2} G_F m_t^2 V^*_{ts} V_{tb}}{16 \pi^2}.
\end{eqnarray}
The finite-width effects of the mediator field $\phi$
are taken into
account by the replacement
in the $\phi$-propagator in Fig.\ref{Fig:1}:
\begin{eqnarray}
  \label{Amp}
 \frac{1}{M_\phi^2-q^2-i\, 0}\to  \frac{1}{M_\phi^2-q^2-i\, M_\phi \Gamma_\phi(q^2)}.
\end{eqnarray}
Equation~(\ref{Amp}) provides a simplified parametrization of the full propagator of the scalar particle,
taking into account the resummation of the $\bar\chi\chi$ loops and neglecting the real parts of the loop
diagrams.\footnote{Taking into account that the correction to the vector-meson propagator due to the pion
loops has precisely the same analytic expression as the correction to the scalar-particle propagator due
to the spin-1/2 fermion loops, one can directly use the real part of the loop diagram given by Eq.~(11) of \cite{gs}.
However, taking the real part into account has a negligible impact on our results,
so we make use of a simplified expression (\ref{Amp}).}
We assume that $M_\phi> 2 m_\chi$ and furthermore that the mediator $\phi$ decays predominantly into the $\bar\chi\chi$ pair. As a consequence, we obtain a $q^2$-dependent width of the mediator $\phi$ which is calculated from the imaginary part of the fermion-loop diagram with scalar vertices in the form (cf. \cite{gs,nachtmann}) 
\begin{equation}
\label{Gamma}
\Gamma_\phi(q^2)= \left(\frac{q^2-4m_\chi^2}{M_\phi^2-4m_{\chi}^2}\right)^{\frac{3}{2}}\frac{M_\phi}{\sqrt{q^2}}\; \Theta(q^2-4m_\chi^2)\; \Gamma_\phi^0, \qquad \Gamma_\phi^0=\frac{g^2_{\phi\chi\chi}}{8\pi}M_\phi\left(1-\frac{4m_\chi^2}{M_\phi^2}\right)^{\frac{3}{2}}.
\end{equation}
Here, $q$ is the momentum of the outgoing $\chi\bar\chi$ pair of unobserved DM particles; $M_X^2\equiv q^2$ is the missing mass squared. 

Using the QCD equations of motion, the required amplitudes can be straightforwardly calculated, yielding
\begin{eqnarray}
\label{FF}
 \langle K|\bar s_L b_R|B \rangle &=&\frac{1}{2}\langle K|\bar s (1-\gamma_5) b|B\rangle=
 \frac{1}{2}\langle K|\bar s b|B\rangle=
 \frac{1}{2} \frac{M_B^2-M_K^2}{m_b-m_s}f_0^{B\to K}(q^2),\nonumber\\
 \langle K^*|\bar s_L b_R|B \rangle &=& \frac{1}{2}\langle K^*|\bar s(1-\gamma_5) b|B \rangle=
 -\frac{1}{2}\langle K^*|\bar s \gamma_5 b|B \rangle
 =-i (\epsilon q)\frac{M_{K^*}}{m_b+m_s}A_0^{B\to K^*}(q^2),
\end{eqnarray}
with well-known dimensionless form factors $f_0^{B\to K}$ and $A_0^{B\to K^*}$ parametrizing
the amplitudes $\langle K|\bar s \gamma_\mu b|B\rangle$ and $\langle K^*|\bar s \gamma_\mu\gamma_5 b|B\rangle$ \cite{wsb}.
For the differential distributions one obtains the following expressions
(we correct here a factor 2 which was missed in the right-hand sides of Eqs.~(2.9) and (2.11) in \cite{blm2025}): 
\begin{eqnarray}
\frac{d\Gamma(B\xrightarrow{\phi} K \bar\chi\chi)}{dq^2}&=&
\frac{\lambda^{1/2}(M_B^2,M_K^2,q^2)}{128 \pi^3 M_B^3}
\frac{(M_B^2-M_K^2)^2 |f_0^{B\to K}(q^2)|^2}{4(m_b-m_s)^2}
\frac{g_{bs\phi}^2\,g_{\phi\chi\chi}^2}{(M_\phi^2-q^2)^2+M_\phi^2\Gamma_\phi^2(q^2)}
 q^2 \left (1-\frac{4 m_\chi^2}{q^2}\right)^{3/2},\nonumber
\label{eq:dGdq2}\\
\frac{d\Gamma(B\xrightarrow{\phi} K^* \bar\chi\chi)}{dq^2}
&=&\frac{\lambda^{3/2}(M_B^2,M_{K^*}^2,q^2)}{128 \pi^3 M_B^3} \frac{|A_0^{B\to K^*}(q^2)|^2}{4(m_b+m_s)^2}
 \frac{g_{bs\phi}^2\,g_{\phi\chi\chi}^2}{(M_\phi^2-q^2)^2+M_\phi^2\Gamma_\phi^2(q^2)}
 q^2 \left(1-\frac{4 m_\chi^2}{q^2}\right)^{3/2}, 
\end{eqnarray}
where $\lambda(a,b,c)=(a-b-c)^2-4bc$.

A powerful probe of the DM scenario considered is provided by the ratio \cite{Berezhnoy:2023rxx}
\begin{eqnarray}
\label{R}
\hat R^{(\phi)}_{K^*/K}(q^2)&=&\frac{d\Gamma(B\xrightarrow{\phi} K^*\bar\chi\chi)/dq^2}{d\Gamma(B\xrightarrow{\phi} K\bar\chi\chi)/dq^2}=
\frac{\lambda^{3/2}(M_B^2,M_{K^*}^2,q^2)}{\lambda^{1/2}(M_B^2,M_K^2,q^2)}
  \frac{|A_0^{B\to K^*}(q^2)|^2}{|f_0^{B\to K}(q^2)|^2}\frac{(m_b-m_s)^2}{(M_B^2-M_K^2)^2 (m_b+m_s)^2}.
\end{eqnarray}
Clearly, the dependence on the specific parameters of the DM model cancels out in this ratio.
On the other hand, as we shall see later, the shape of this function may be used as a clear signature of the mediator spin. 

The form factors $f^{B\to K}_0$ and $A_0^{B\to K^*}$ are treated as external inputs for our analysis and 
are given by nonperturbative QCD calculations. The parametrizations of the form factors $f^{B\to K}_0(q^2)$
and $A_0^{B\to K^*}(q^2)$ will be discussed in Sec.~\ref{Sect:3}, where numerical results for the
observables are presented.

\subsection{V-scenario}
One of the widely discussed top-philic vector-mediator scenario is described by the Lagrangian
\cite{Langacker:2008yv,Cox:2015afa,Hu:2024xes}
\begin{eqnarray}
\mathcal{L}_{\rm int}=g_{Vtt}\,V^\mu\,\bar t \gamma_\mu (1+\gamma_5)t+g_{V\chi\chi}\, V^\mu\,\bar\chi \gamma_\mu \chi. \nonumber
\end{eqnarray}
Coupling of the mediator $V$ to the FCNC $b\to s$ current occurs through the penguin loop diagram of Fig.~1. 
Integrating out the heavy top and $W$ fields leads to the $b\to sV$ effective Lagrangian which has
the $V-A$ structure \cite{Inami:1980fz}
\begin{eqnarray}
\label{LbsV} 
\mathcal{L}_{b \rightarrow sV}=g_{bs V} \bar s\gamma_\mu (1-\gamma_5)b\, V^\mu. 
\end{eqnarray}
The Wilson coefficient $g_{bs V}$ is obtained from the triangle diagram of Fig.~1 with the
$\bar t_R\gamma_\mu t_R$ current in one of the vertices and the SM $(V-A)$ currents in two other vertices. 
The necessary result for $g_{b s V}$ may be obtained using Eq.~(2.6) of Ref.~\cite{Inami:1980fz}
for the sum of two diagrams, $\Gamma^{(c)}$ and $\Gamma^{(d)}$ (see Eq.~(A.1) from \cite{Inami:1980fz}).
Furthermore, one should set in these expressions $\cos\,\theta_{\rm W}\to 1/2$,
$\sin\,\theta_{\rm W}\to \sqrt{3}/2$, 
and replace $g^3\to -8\sqrt{2}g_{ttV}G_F M_W^2$.
Using dimensional regularization, the diagrams $\Gamma^{(c)}$ and $\Gamma^{(d)}$
contain $1/\epsilon$ divergent terms (unlike the SM calculation, we have only $t$-quarks in the loop and so
the GIM mechanism does not work). In the $\overline{\rm MS}$ renormalization scheme, we
come to the following expression [$g_{bsV}\equiv \bar g_{bsV}(M_W)$]: 
\begin{eqnarray}
  \label{gbsV}
 g_{bsV}&=&
 -\frac{g_{Vtt}}{16\pi^2}\frac{G_F}{\sqrt{2}}V_{tb}V_{ts}^*\,m_t^2\,F(x_t),\quad x_t=(m_t/M_W)^2,\nonumber \\
 &&F(x_t)=\frac{7-8x_t+x_t^2+2(4+(x_t-2)x_t)\log\,x_t}{(1-x_t)^2}=0.19.
\end{eqnarray}
To obtain the differential distributions $d\Gamma(B\xrightarrow{V} h\bar\chi\chi)/dq^2$, $h=K,K^*$,
we may use the results of \cite{Melikhov:1997wp}, Eqs.~(32) and (36), and replace there
$C_{7\gamma}\to 0$, $C_{10A}\to 0$, and
\begin{eqnarray}
\frac{G_F}{\sqrt{2}}\frac{\alpha_{\rm em}}{2\pi}V_{tb}V_{ts}^*C_{9V}\to
  g_{bsV}\frac{1}{M_V^2-q^2-i\Gamma_V(q^2) M_V}g_{V\chi\chi}. 
\end{eqnarray}
In the end, we obtain ($h=K,K^*$): 
\begin{eqnarray}
  \frac{d\Gamma(B\xrightarrow{V}h\bar\chi\chi)}{dq^2}=
  \frac{M_B^3}{192\pi^3}\sqrt{\lambda(1,\hat q^2,r_h^2)}\left(1+\frac{2m_\chi^2}{q^2}\right)\sqrt{1-\frac{4m_\chi^2}{q^2}}
  \frac{g_{bsV}^2 g_{V\chi\chi}^2}{(M_V^2-q^2)^2+\Gamma^2_V(q^2)M_V^2}\,\beta_h , 
\end{eqnarray}
with $\hat q^2=q^2/M_B^2$, $r_h=M_h/M_B$, and 
\begin{eqnarray}
  \beta_K&=& |f_+^{B\to K}(q^2)|^2\lambda(1,\hat q^2,r_K^2)  \\
  \beta_{K^*}&=&\frac{1}{4M_B^4M_{K^*}^2}
  \left| A^{B\to K^*}_2(q^2) \frac{\lambda(q^2,M_B^2,M^2_{K^*})}{M_B+M_{K^*}}-A^{B\to K^*}_1(q^2)(M_B^2-M_{K^*}^2-q^2)(M_B+M_{K^*})\right|^2\nonumber\\
  &&+\frac{2|A^{B\to K^*}_1(q^2)|^2(M_B+M_{K^*})^2q^2}{M_B^4}+\frac{2\lambda(q^2,M_B^2,M^2_{K^*})q^2 |V^{B\to K^*}(q^2)|^2}{M_B^4(M_B+M_{K^*})^2}.
\end{eqnarray}
Assuming that $M_V>2m_\chi$, the $q^2$-dependent width $\Gamma_V(q^2)$ is obtained as the imaginary part of the
two-point diagram with fermions in the loop and vector vertices \cite{nachtmann} leading to 
\begin{eqnarray}
  \label{GammaV}
  \Gamma_V(q^2)=\frac{M_V}{\sqrt{q^2}}
  \sqrt{\frac{q^2-4m_\chi^2}{M_V^2-4m_{\chi}^2}}\frac{q^2+2m_\chi^2}{M_V^2+2m_\chi^2}\Theta(q^2-4m_\chi^2)\; \Gamma_V^0, 
\quad \Gamma_V^0=\frac{g^2_{V\chi\chi}M_V}{12\pi}\sqrt{1-\frac{4m_\chi^2}{M_V^2}}\left(1+\frac{2m_\chi^2}{M_V^2}\right).
\end{eqnarray}
These expressions contain the DM parameters $M_V$, $m_\chi$, $g_{bsV}$, and $g_{V\chi\chi}$ which may be extracted by fitting the data. 

A valuable probe of the DM scenario is provided by the ratio
\begin{eqnarray}
\hat R^{(V)}_{K^*/K}(q^2)=\frac{d\Gamma(B\xrightarrow{V} K^*\bar\chi\chi)/dq^2}{d\Gamma(B\xrightarrow{V}K\bar\chi\chi)/dq^2}
\end{eqnarray}
This ratio is not sensitive to the specific values of the DM parameters (which cancel in the ratio),
but is very sensitive to the mediator spin, as we shall demonstrate in Section \ref{Sect:3}. 

The form factors $f^{B\to K}_+(q^2)$, $A_{1,2}^{B\to K^*}(q^2)$ and $V^{B\to K^*}(q^2)$ are standard quantities
defined in \cite{wsb}. These form factors are obtained using full QCD including nonperturbative QCD effects. 
For our discussion the form factors are external inputs; we shall use the parametrizations known in the literature.
The adopted parametrizations will be discussed in Section \ref{Sect:3}.

In what follows, we are going to address two partly independent problems: 

(i) Work out signatures for S- and V-scenarios

(ii) Determine the DM parameters for S- and V-scenarios by fits to the available data.

\section{\label{Sect:3} Signatures of S- and V-scenarios}

\subsection{The $B\to (K,K^*)$ form factors} 
For the $B\to K$ form factors $f_0$ and $f_+$ we make use of the results from lattice QCD \cite{Bailey:2015dka}
in the~form of rather convenient parametrizations proposed in \cite{ms2000}:
\begin{eqnarray}
\label{fplus}
f^{B\to K}_+(q^2) &=& \frac{0.335}{(1-q^2/M_{RV}^2)(1-0.58\,q^2/M_{RV}^2+0.03 (q^2/M_{RV}^2)^2)}, \\
\label{f0}
f^{B\to K}_0(q^2) &=& \frac{0.335}{1-0.648\,q^2/M_{RV}^2-0.17 (q^2/M_{RV}^2)^2},
\qquad M_{RV}=M_{B_s}(1^-)=5.415\mbox{ GeV}.
\end{eqnarray}
These form factors are not far from other determinations of these quantities.
An uncertainty of 10\% may be safely assigned to $B\to K$ form factors.

For the $B\to K^*$ form factors $A^{B\to K^*}_{0,1,2}$ and $V^{B\to K^*}$ the situation is more complicated. This is related to the necessity to isolate the vector meson in the final state.
We make use of the results from light-cone sum rules 
for the form factors $A_0$ and $V$ \cite{Bharucha:2015bzk} and $A_{1,2}$ \cite{Khodjamirian:2010vf}, 
again recalculated to the parametrization of \cite{ms2000}:  
\begin{eqnarray}
\label{V}
V^{B\to K^*}(q^2) &=& \frac{0.36}{(1-q^2/M_{RV}^2)(1+0.15 q^2/M_{RV}^2 +0.166(q^2/M_{RV}^2)^2)},
  \qquad M_{RV}=M_{B_s}(1^-)=5.415\mbox{ GeV},\\
\label{A0}
A^{B\to K^*}_0(q^2) &=& \frac{0.37}{(1-q^2/M_{RS}^2)(1-0.656 q^2/M_{RS}^2 +0.03(q^2/M_{RS}^2)^2)},
  \qquad M_{RS}=M_{B_s}(0^-)=5.366\mbox{ GeV},\\
\label{A1}
A^{B\to K^*}_1(q^2) &=& \frac{0.25}{(1-q^2/M_{RA}^2)(1+0.097 q^2/M_{RA}^2 +0.02(q^2/M_{RA}^2)^2)},
\qquad M_{RA}=M_{B_s}(1^+)=5.829\mbox{ GeV},\\
\label{A2}
A^{B\to K^*}_2(q^2) &=& \frac{0.23}{(1-q^2/M_{RA}^2)(1-0.114 q^2/M_{RA}^2 -0.128(q^2/M_{RA}^2)^2)}.
\end{eqnarray}
Taking account of the scattered results from various nonperturbative methods (in some cases well beyond the
estimated uncertainties), it seems reasonable to assign the uncertainty of at least 15\% to the $B\to K^*$ form factors. 
\subsection{The ratio of the differential rates in $B\xrightarrow{R} K^*\bar\chi\chi$ over $B\xrightarrow{R}K\bar\chi\chi$}
Let us consider the ratio of the differential distributions of the excess events in S- and V-scenarios: 
\begin{eqnarray}
  \label{ratio_diff}
  \hat R^{(R)}_{K^*/K}(q^2)=\frac{d\Gamma(B\xrightarrow{R}K^*\bar\chi\chi)/dq^2}{d\Gamma(B\xrightarrow{R} K\bar\chi\chi)/dq^2},
  \quad R=\phi, V. 
\end{eqnarray}
In each of these scenarios, this ratio is insensitive to DM parameters
($M_R$, $m_\chi$, $g_{bsR}$, $g_{R\chi\chi}$) as they cancel in the ratio.
Thus, the latter may be calculated unambiguously for each of the DM scenarios and
exhibits a visible dependence on the mediator spin.
\begin{center}
\begin{figure}[h]
\begin{tabular}{cc}
\includegraphics[width=6.5cm]{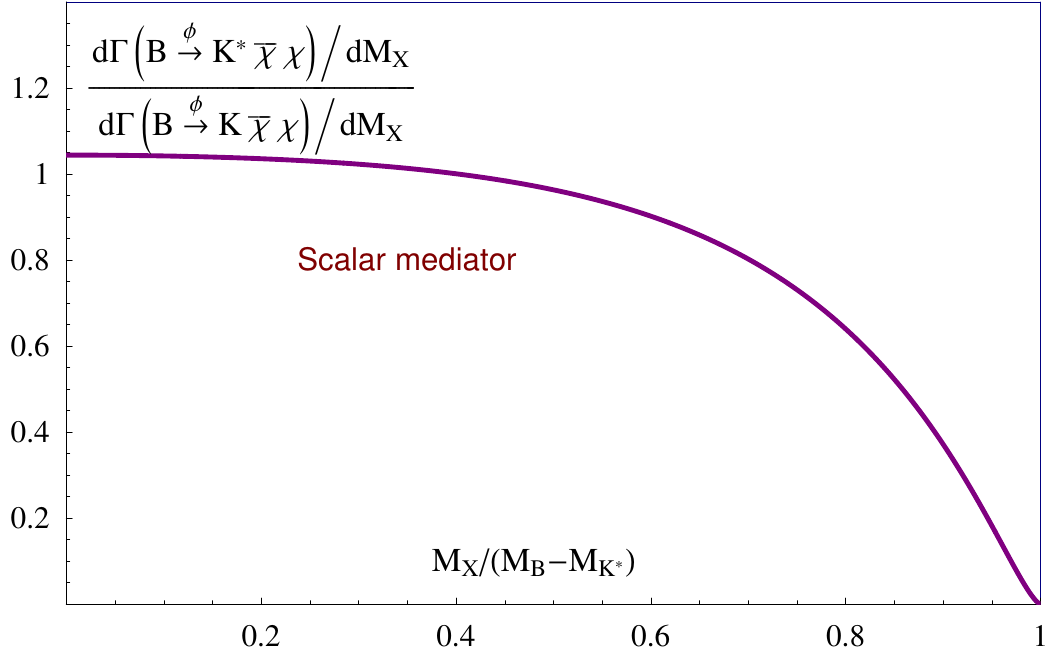} \hspace{.5cm} & \hspace{.5cm}
\includegraphics[width=6.5cm]{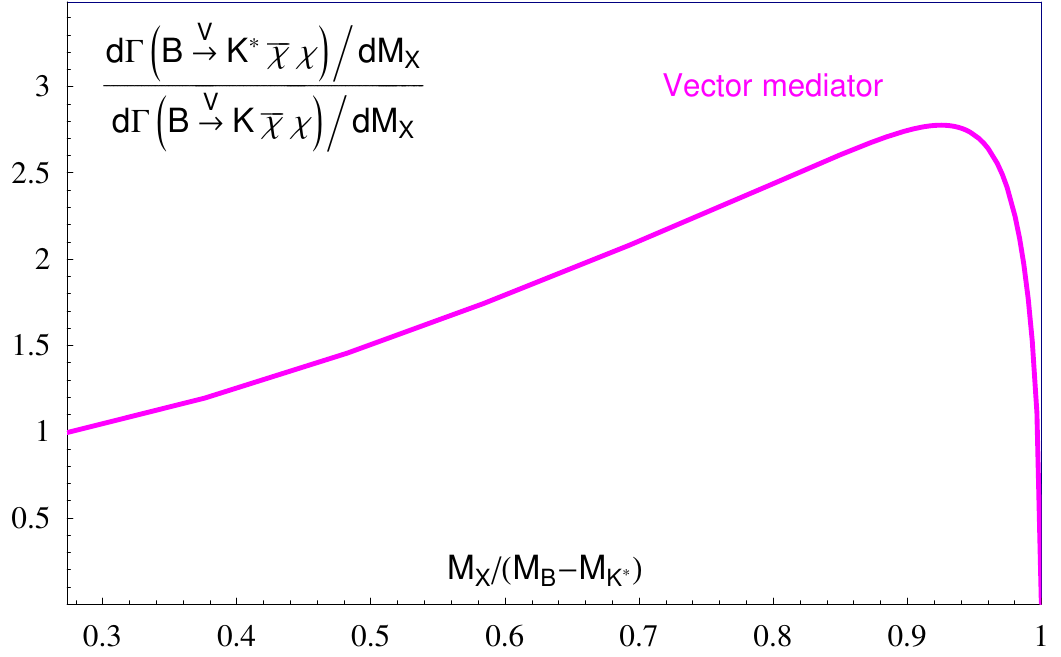}
\\
(a)   &   (b)
\end{tabular}
\caption{\label{Fig:2} The ratio of the differential distributions in
  $B\to K^*\bar\chi\chi$ over $B\to K\bar\chi\chi$ for S-scenario (a) and V-scenario (b).}
\end{figure}
\end{center}
Figure~\ref{Fig:2} shows the results of calculations using formulas of Sec.~\ref{Sect:2}
and the $B\to (K,K^*)$ form factors described above. Clearly, the ratio  $\hat R^{(R)}_{K^*/K}(q^2)$
has qualitatively different behaviors for S- and V-scenarios: Within the S-scenario, the ratio decreases with $q^2$ and stays
less than 1 in a broad region of $q^2$.  Within the V-scenario, the ratio rises with $q^2$
and stays above 1 in almost the
full kinematically available region of $q^2$.

The ratio $\hat R^{(R)}_{K^*/K}(q^2)$ may be measured experimentally, thus providing a clear signature of the DM scenario
(i.e., the mediator spin), independently of DM parameters. The form factor uncertainties mentioned above
cancel to a large extent in the ratio, so the uncertainty in our predictions for this ratio does not exceed a 5\% level. 
\subsection{The ratio of the integrated rates in $B\xrightarrow{R} K^*\bar\chi\chi$ over $B\xrightarrow{R} K\bar\chi\chi$}
\begin{center}
\begin{figure}[h]
\begin{tabular}{cc}
\includegraphics[width=6.5cm]{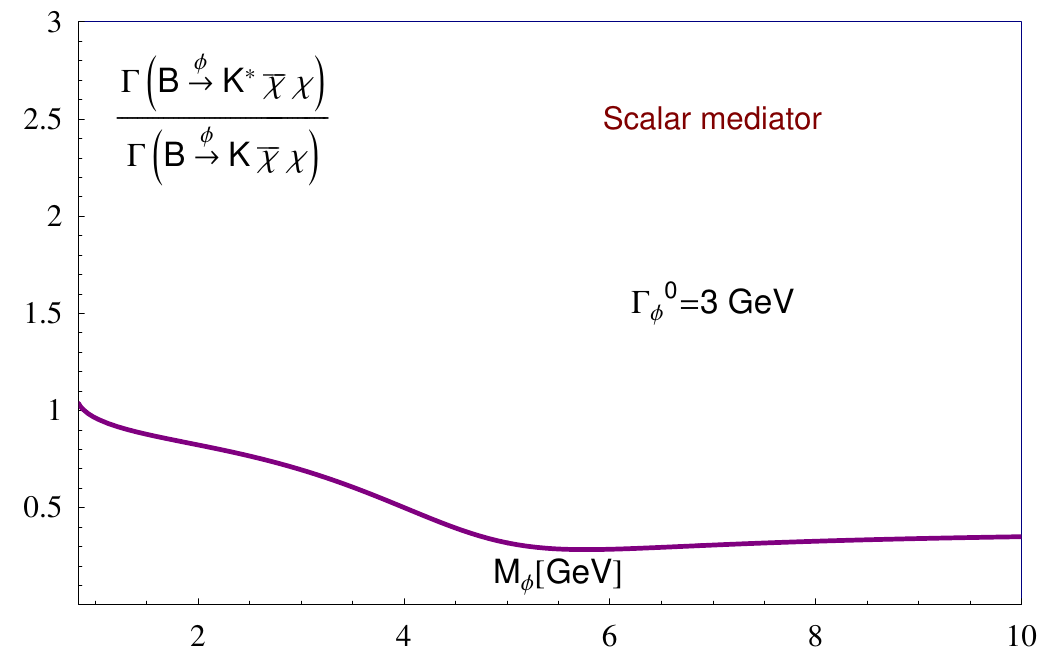}\hspace{.5cm} & \hspace{.5cm}
\includegraphics[width=6.5cm]{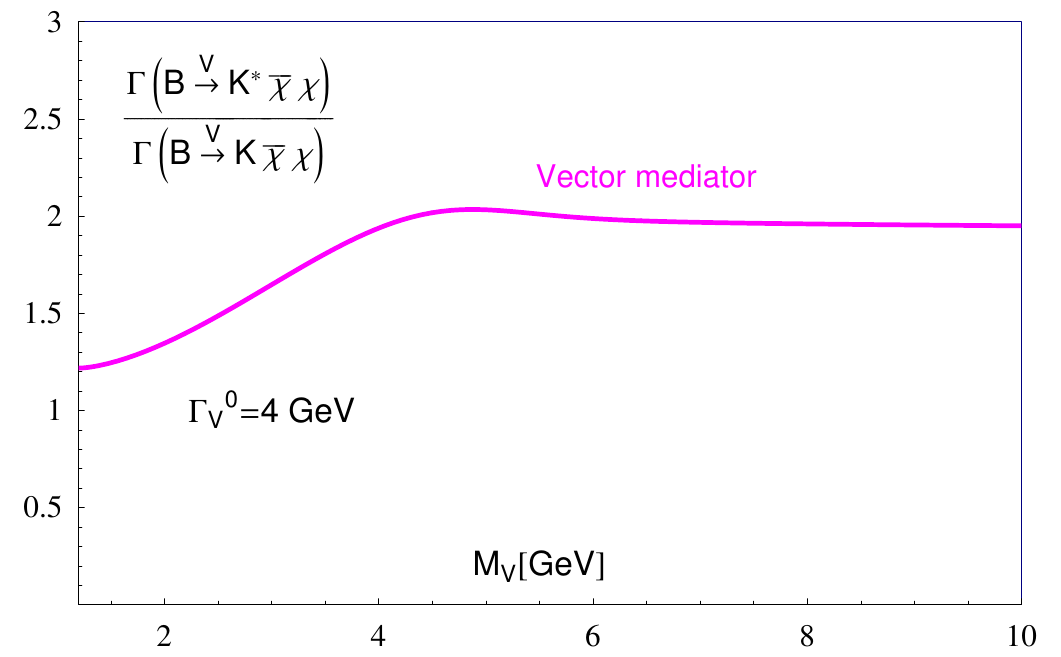}
\\
(a) & (b)
\\
\end{tabular}
\caption{\label{Fig:3} The ratio of the integrated rates $R^{(R)}_{K^*/K}$, Eq.~(\ref{ratio_int}),
  for S-scenario (a) and V-scenario (b).}
\end{figure}
\end{center}

We now consider the ratio of the {\it integrated} rates of the excess events
\begin{eqnarray}
\label{ratio_int}
R^{(R)}_{K^*/K}=\Gamma(B\xrightarrow{R} K^*\bar\chi\chi)/\Gamma(B\xrightarrow{R}K\bar\chi\chi). 
\end{eqnarray}
This ratio shows a strong sensitivity to the mediator mass $M_R$ and (to a lesser extent) to the mediator width $\Gamma_0^R$.
Crucial for us is that the dependencies on $M_R$ are qualitatively different in S-and V-scenarios, see Fig.~\ref{Fig:3}. 
We emphasize the qualitative difference between the behavior of $R^{(\phi)}_{K^*/K}$ and $R^{(V)}_{K^*/K}$:
independently of the mediator mass, $R^{(\phi)}_{K^*/K}<1$ whereas  $R^{(V)}_{K^*/K}>1$.
Measuring of $R^{(R)}_{K^*/K}$ thus provides an additional valuable probe of the mediator spin. 

\subsection{Constraints on $B\to K^*M_X$ based on the measured $B\to K M_X$}
\begin{center}
\begin{figure}[h]
\begin{tabular}{cc}
  \includegraphics[width=6.5cm]{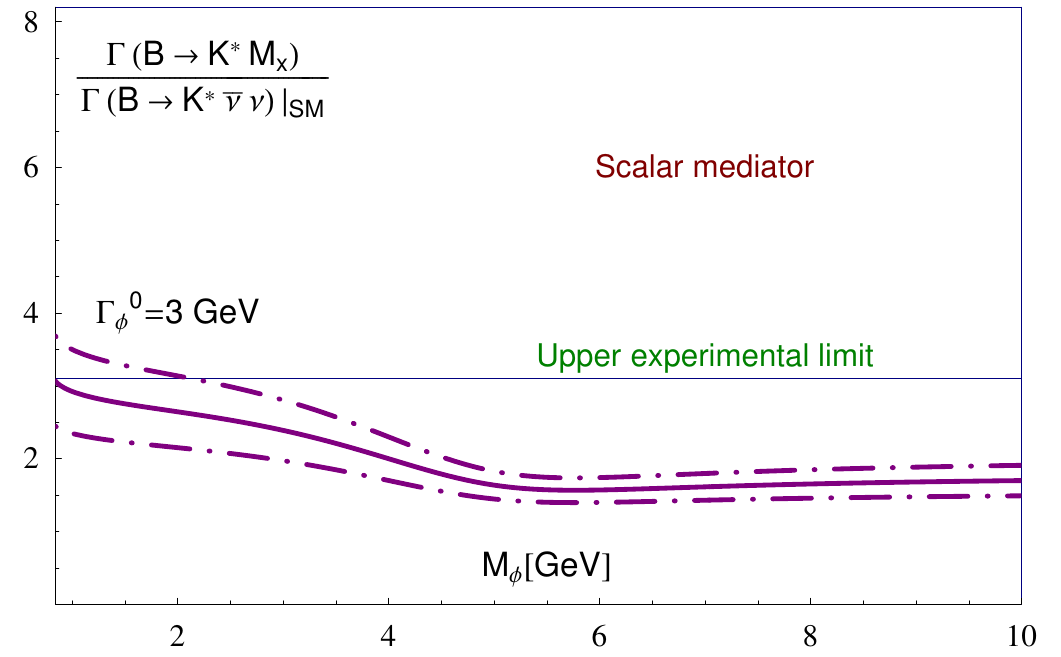} \hspace{.5cm} &
  \hspace{.5cm} \includegraphics[width=6.5cm]{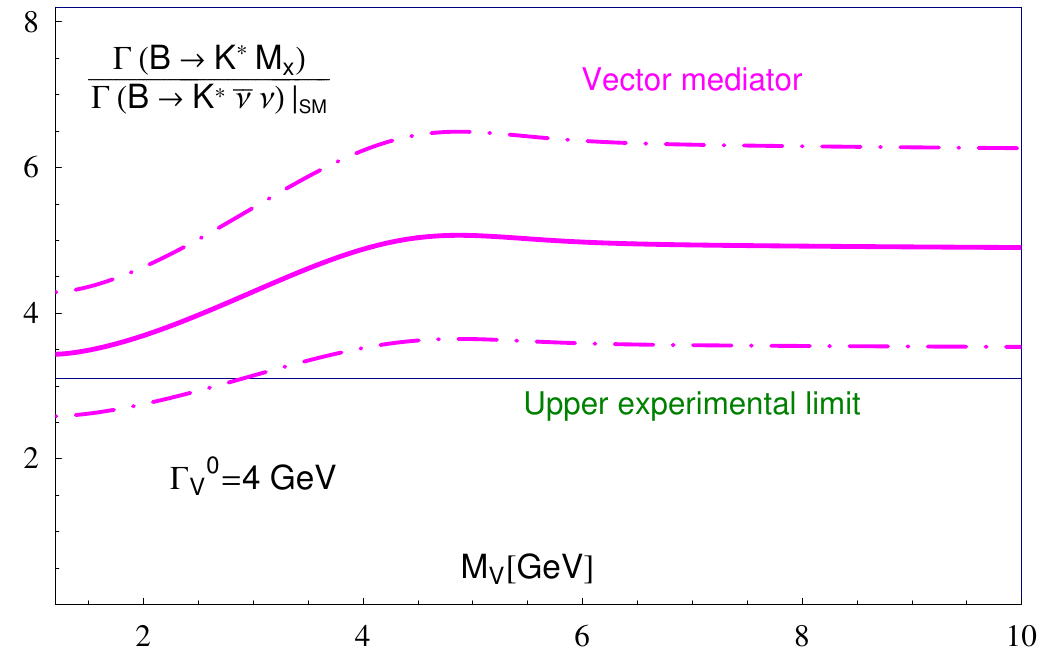}\\
(a) & (b)
\end{tabular}
\caption{\label{Fig:4} The ratio of the integrated rates 
  $\Gamma(B\to K^* M_X)/\Gamma(B\to K^*\bar\nu\nu)_{\rm SM}$ for S-scenario (a) and
  V-scenario (b). The solid line corresponds to $1+2 R^{(R)}_{K^*/K}$, Eq.~(\ref{ratiooverSM}),
  the upper and lower dashed lines correspond to $1+(2\pm 0.6)R^{(R)}_{K^*/K}$.
  The horizontal line shows the upper experimental limit for this ratio, Eq.~(\ref{constraintonMR}).}
\end{figure}
\end{center}

We now combine our results for $R^{(R)}_{K^*/K}$ with the theoretical estimates for the decay rates
$\Gamma(B\to (K,K^*)\bar\nu\nu)_{\rm SM}$ and the conjecture that the observed excess in $B\to K M_X$
is due to the decay into invisible DM fermions. We can write the following sequence of identities 
\begin{eqnarray}
\label{ratioSM}
\frac{\Gamma(B\to K^*M_X)}{\Gamma(B\to K^*\bar\nu\nu)_{\rm SM}}& =&
\frac{\Gamma(B\to K^*\bar\nu\nu)_{\rm SM}+\Gamma(B\to K^*\bar\chi\chi)}{\Gamma(B\to K^*\bar\nu\nu)_{\rm SM}}\nonumber\\
&=&1+\underbrace{\frac{\Gamma(B\to K^*\bar\chi\chi)}{\Gamma(B\to K\bar\chi\chi)}}_{\mbox{Calculated }R_{K^*/K}}
\underbrace{\frac{\Gamma(B\to K\bar\chi\chi)}{\Gamma(B\to K\bar\nu\nu)_{\rm SM}}}_{\mbox{Belle-II: }4.4\pm 1.5}
\underbrace{\frac{\Gamma(B\to K\bar\nu\nu)_{\rm SM}}{\Gamma(B\to K^*\bar\nu\nu)_{\rm SM}}}_{{\rm Theory }:(4.44\pm 0.30)/(9.8\pm 1.4)}.
\end{eqnarray}
Making use of the results for the different factors, indicated in Eq.~(\ref{ratioSM}) with their respective uncertainties,
we obtain
\begin{eqnarray}
\label{ratiooverSM}
\frac{Br(B\to K^*M_X)}{Br(B\to K^*\bar\nu\nu)_{\rm SM}}= 1+(2\pm 0.6)R^{(R)}_{K^*/K}.
\end{eqnarray}
This expression may be combined with the Belle upper limit, Eq.~(\ref{belle2017}), and a theory
estimate $Br(B\to K^*\bar\nu\nu)_{\rm SM}=(9.8\pm 1.4)\cdot 10^{-6}$, Eq.~(\ref{damir}),
yielding the following constraint:
\begin{eqnarray}
  \label{constraintonMR}
\frac{Br(B\to K^*M_X)}{Br(B\to K^*\bar\nu\nu)_{\rm SM}}<3.1. 
\end{eqnarray}
Fig.~\ref{Fig:4} displays the ratio (\ref{ratioSM}) for the S- and V-scenarios.
Clearly, the condition (\ref{constraintonMR}) is fulfilled for any value of the scalar mediator mass $M_\phi$.
So, within S-scenario no restrictions on the mediator mass emerge from the upper experimental limit Eq.~(\ref{constraintonMR}). 
However, within the V-scenario Eq.~(\ref{constraintonMR}) yields a tight constraint on $M_V$ --
only a light vector mediator is allowed,
\begin{eqnarray}
  \label{MV}
M_V\lesssim 3 \mbox{ GeV}. 
\end{eqnarray}
We shall use this constraint when fitting the Belle-II results for the
differential distributions in $B\to KM_X$. 

\section{\label{Sect:4} BELLE-II data fitting and extraction of DM parameters}
The Belle-II experiment, as described in \citep{Belle-II:2023esi},
analyzes partially reconstructed events to increase the statistical sample.
For such events, the direction of the $B$ meson cannot be determined.
Therefore, instead of $q^2=(p_B-p_K)^2$ the variable $q_{\rm rec}^2$ is used:
 \begin{equation}
q^2_{\rm rec}= E_B^2+M_K^2-2E_B E_K,
 \end{equation}
 where $E_B$ and $E_K$ are the energies of the $B$ and $K$ mesons in the
 center-of-mass frame of the $B\bar B$-meson pair produced in $\Upsilon(4S)$ decays.

 Our theoretical predictions for the differential distributions are obtained as 
 functions of $q^2$. In order to apply them to the analysis of the Belle-II data given as functions of $q_{\rm rec}^2$,
 an appropriate recalculation (averaging) of our results $q^2\to q_{\rm rec}^2$ should be done. 
 The recalculation procedure as well as the data fitting algorithm that yields the DM parameters 
 were described in full detail in Sections III and IV of \cite{blm2025} for the S-scenario and will not be repeated here.
 We now apply precisely the same procedures for the V-scenario.



 \subsection{Fits to the Belle-II differential distribution}
\begin{figure}[t]
\begin{tabular}{ccc}
  \includegraphics[width=8cm]{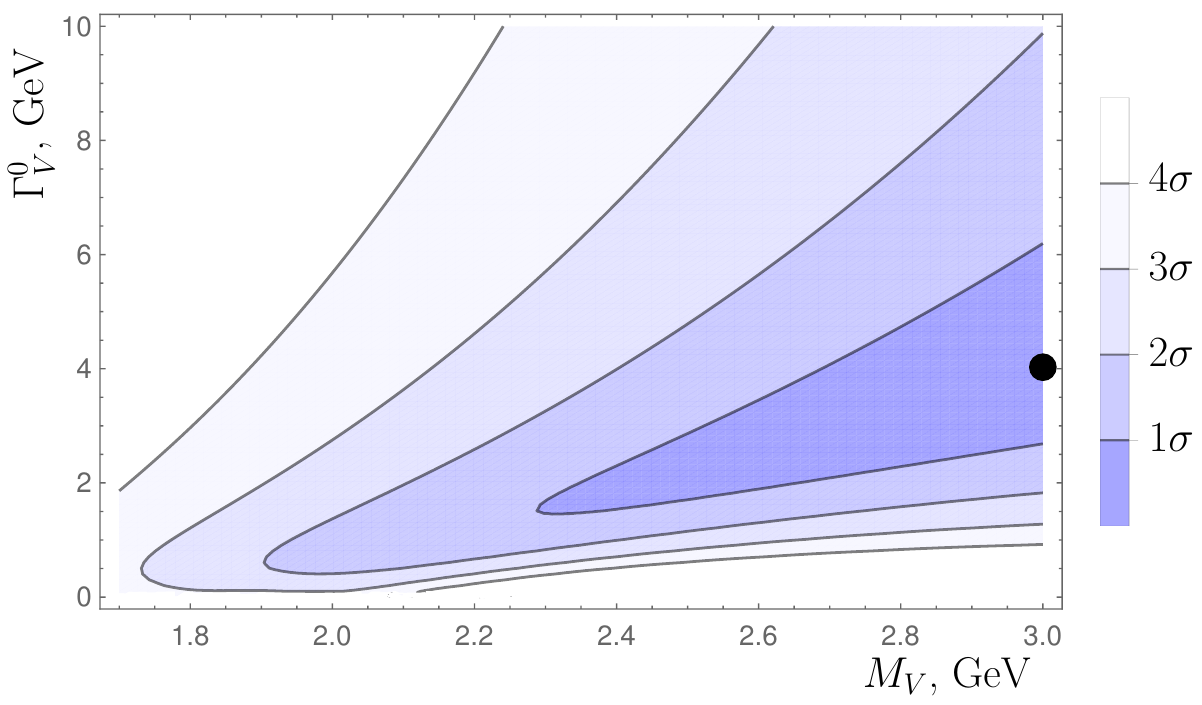} & ${}$\hspace{.5cm} ${}$ &
  \includegraphics[width=8cm]{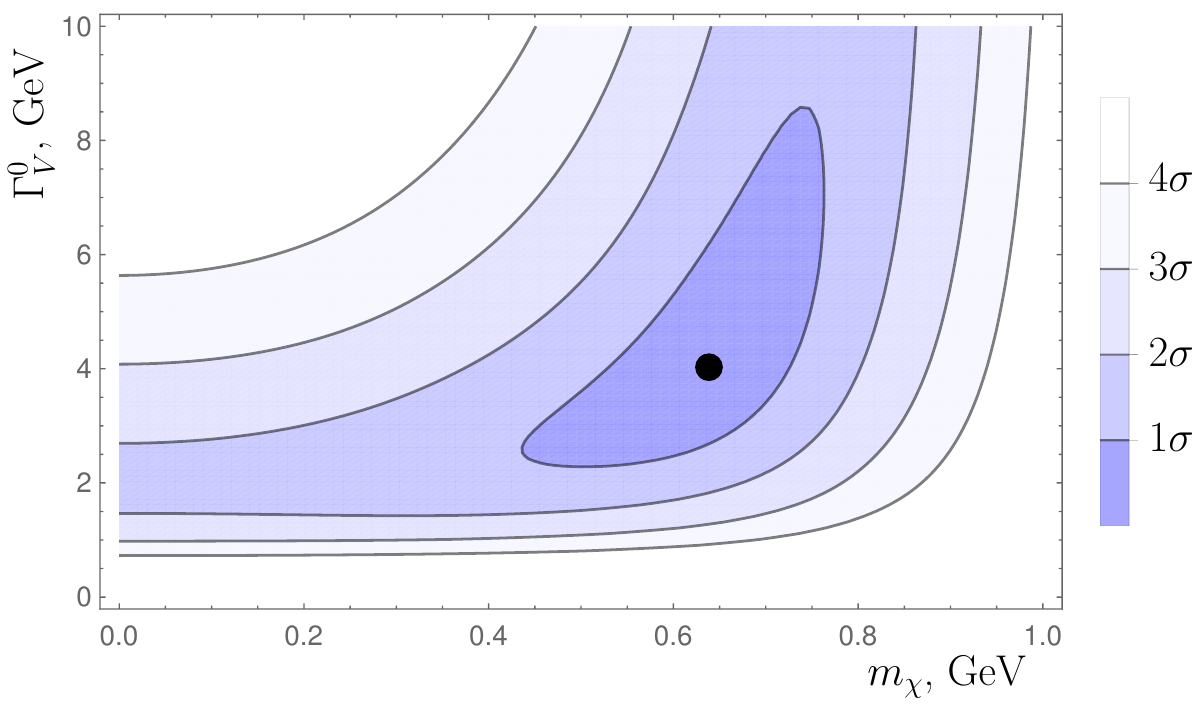}\\
(a) & & (b) \vspace{.5cm}\\
\includegraphics[width=8cm]{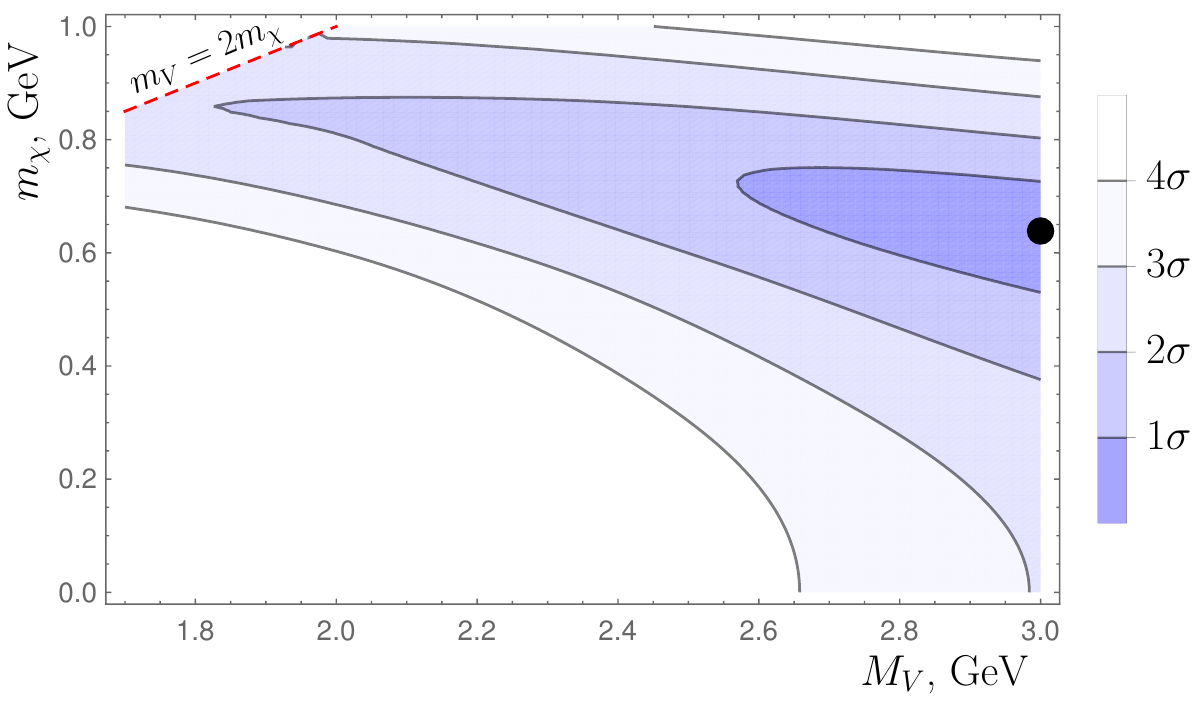} & & 
\includegraphics[width=7.2cm]{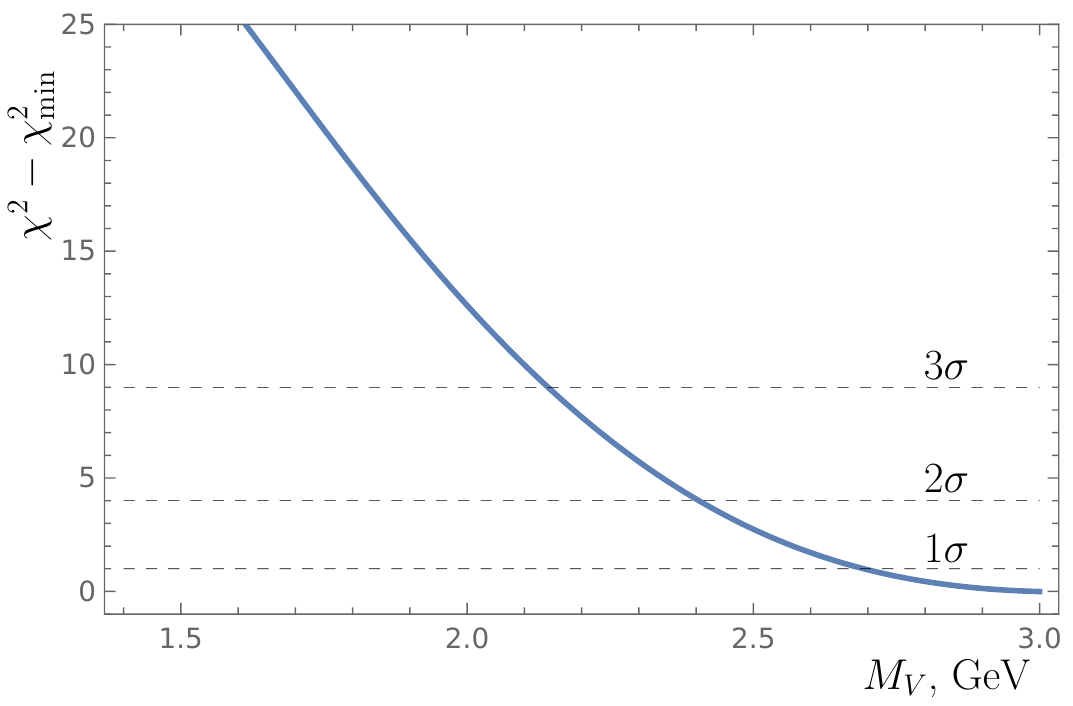}\\
(c) & & (d)\\
\includegraphics[width=7.2cm]{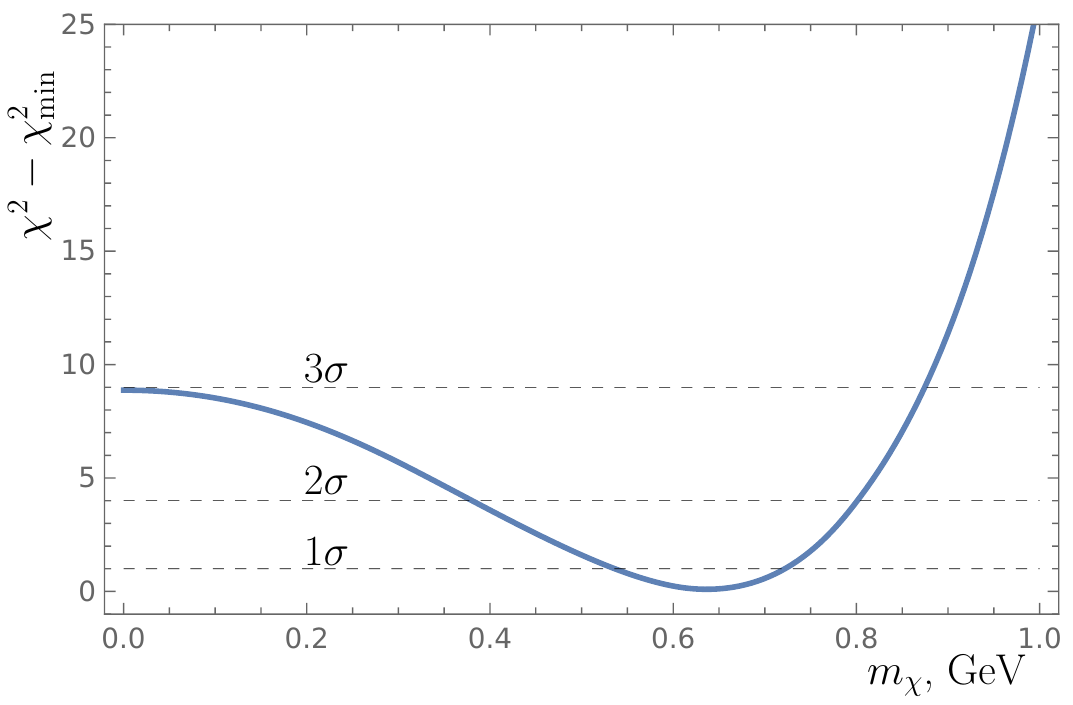} & &
\includegraphics[width=7.2cm]{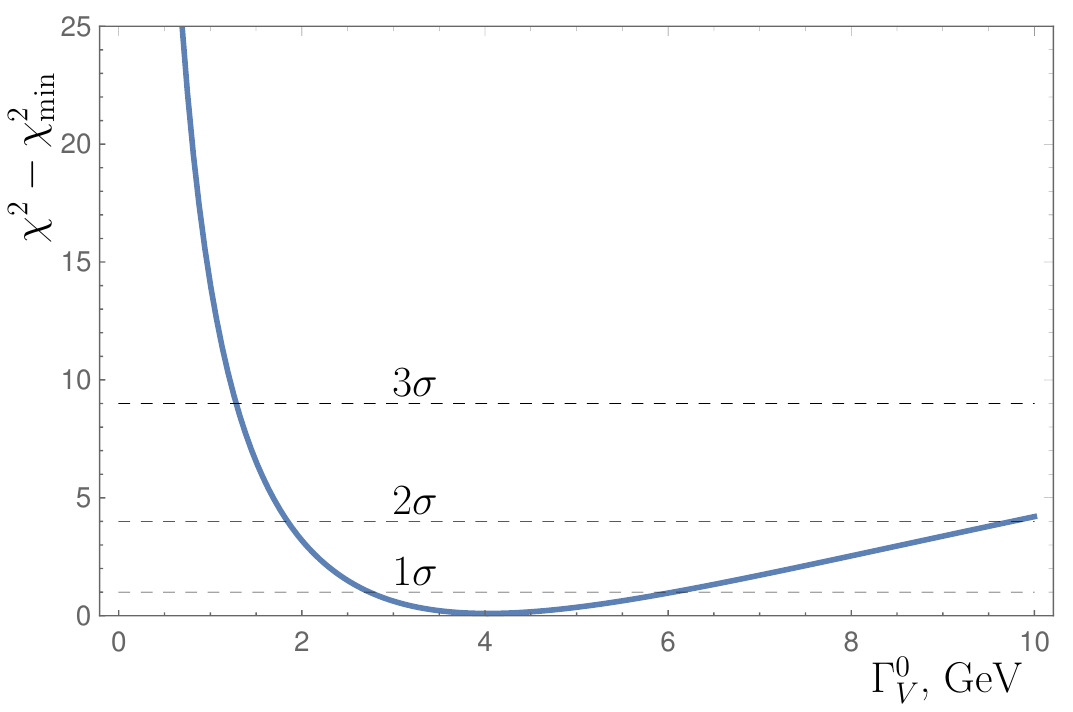}\\
(e) & & (f)\\
\end{tabular}
\caption{The $\chi^2$ distributions. 
(a,b,c) - the two-dimensional (2D) $\chi^2$ distributions
(a) $M_{V}$ and $\Gamma^0_V$,
(b) $m_{\chi}$ and $\Gamma^0_V$,
(c) $M_{V}$ and $m_\chi$. 
The black dots indicate the used values of the parameters corresponding to the minimal 
$\chi^2_{\rm min}=9.04$: $M_\phi=2.4$ GeV, $\Gamma_\phi^0=2.9$ GeV and $m_\chi=0.42$ GeV.
The value of the third variable is set to its ``best'' value. 
(d,e,f) - the one-dimensional (1D) distributions of $\chi^2$ vs
(d): $M_V$, (e): $m_\chi$, (f): $\Gamma_V^0$. The other two parameters in these plots are set to their best values. }
\label{fig_V_chi2}
\end{figure}

 The results of fitting the experimental differential distributions  $d\Gamma/dM_X^2$ by our formulas
 for $V$-scenario are shown in Fig.~\ref{fig_V_chi2}. The two- and one-dimensional distributions of $\chi^2$ are presented.
The corresponding plots for $S$-scenario are given in Fig:~1 of \cite{blm2025}. 

\begin{center}
\begin{figure}[t]
\begin{tabular}{cc}
\includegraphics[width=6.5cm]{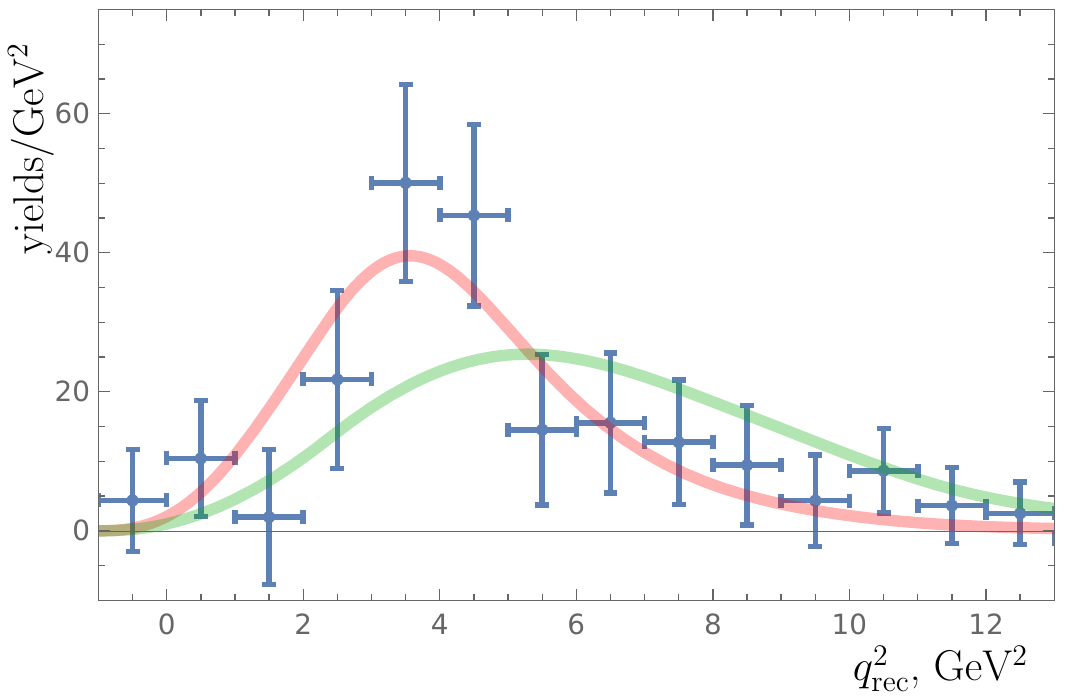}\hspace{.5cm} & \hspace{.5cm}
\includegraphics[width=6.5cm]{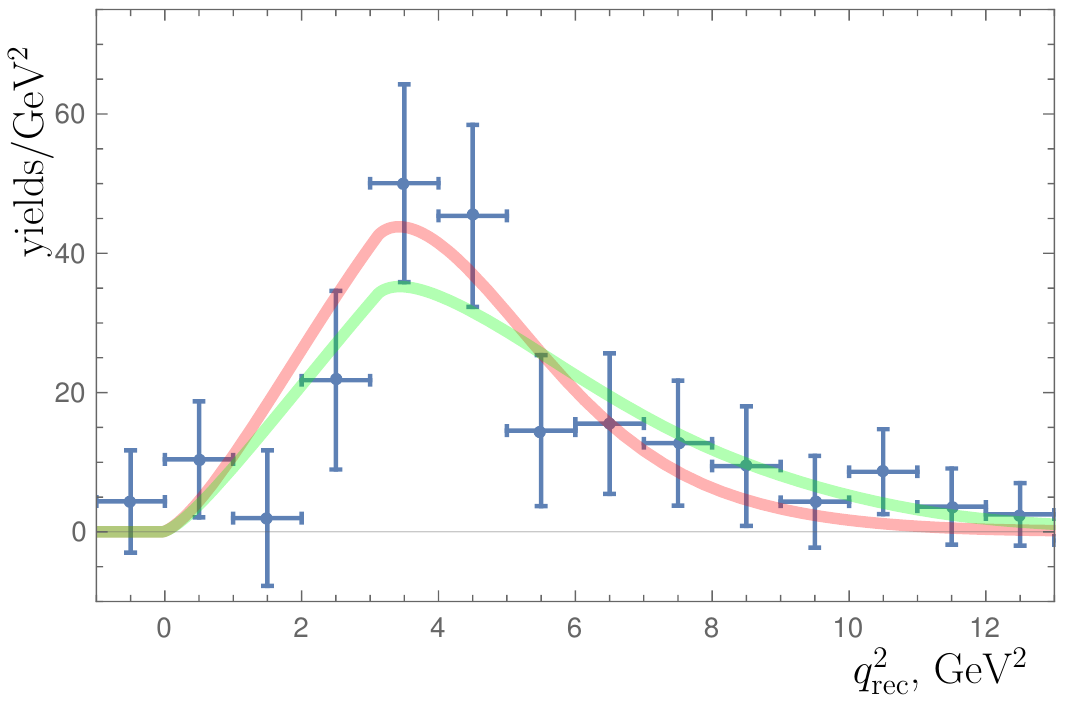}\\
(a)  &  (b)
\end{tabular}
\caption{
  \label{Fig:5}
  The Belle-II data on the decay $B\to K M_X$ (see Fig.~18 from \cite{Belle-II:2023esi} and Fig.~1 from 
  \cite{Fridell:2023ssf}) fitted by our DM models. 
  (a): S-scenario. 
  The ``best'' point corresponding to the global minimum of $\chi^2$,
  $M_\phi=2.4$~GeV, $\Gamma^0_\phi=2.9$~GeV  and $m_\chi=0.42$~GeV (red solid curve);
  a representative point at large $M_\phi=20$~GeV, $\Gamma^0_\phi=20$~GeV and $m_\chi=0.42$~GeV (green solid curve).
(b): V-scenario. The ``best'' point corresponds to the boundary of the allowed region $M_V=3$ GeV, 
$\Gamma_V^0=4$ GeV, $m_\chi=0.6$ GeV (red solid curve);  
a representative point from the region excluded by (\ref{MV}): 
$M_V=20$ GeV, $\Gamma_V^0=4$ GeV, $m_\chi=0.6$ GeV (green solid curve).
}
\end{figure}
\end{center}

Figure~\ref{Fig:5} presents the results of the fits to the Belle-II data in two scenarios.
For the S-scenario, the ``best'' values corresponding to the global minimum of $\chi^2$, were reported in \cite{blm2025}: 
\begin{eqnarray}
  \label{DMS}
  M_\phi=2.4\pm 0.4 \mbox{ GeV},\quad  \Gamma^0_\phi=2.9^{+1.1}_{-0.9} \mbox{ GeV}, \quad  m_\chi=0.42^{+0.2}_{-0.4}\mbox{ GeV}.
\end{eqnarray}
The coupling $g_{\phi\chi\chi}$ is obtained from the extracted values of $M_\phi$ and $\Gamma_\phi^0$
via Eq.~(\ref{Gamma}), $g_{\phi\chi\chi}\approx 6$. From the total number of $B$ mesons produced
at Belle-II ($N_\mathrm{tot}=3.99 \cdot 10^8$), using this $g_{\phi\chi\chi}$ and the observed
excess of approximately 170 events, yields $g_{bs\phi}\sim 5 \cdot 10^{-8}$
[this number is corrected compared to \cite{blm2025}, where a factor 2 was missing in Eq.~(2.11)]. 

For the V-scenario, the situation is a bit more complicated: the best $\chi^2$ is reached on the boundary of the
allowed region [recall that the region $M_V\gtrsim 3$ GeV is excluded by the upper limit on $\Gamma(B\to K^*\bar\nu\nu)$]
so we set $M_V=3$ GeV and determine the other parameters by standard $\chi^2$-criteria: 
\begin{eqnarray}
  \label{DMV}
  M_V=3 \mbox{ GeV},\quad  \Gamma^0_V=4.0^{+2.0}_{-1.5} \mbox{ GeV}, \quad  m_\chi=0.6^{+0.10}_{-0.18}\mbox{ GeV}, 
\end{eqnarray}
and the related couplings $g_{V\chi\chi}\sim 7$ and $g_{bsV}\sim 2\cdot 10^{-8}$.

Recall once more, that without taking into account the constraint (\ref{MV}), nearly any value of $M_V\ge 3$ GeV
yields a reasonable description of the data. 

\vspace{-1cm}

\begin{center}
\begin{figure}[b]
  \begin{tabular}{cc}
    \includegraphics[width=6.5cm]{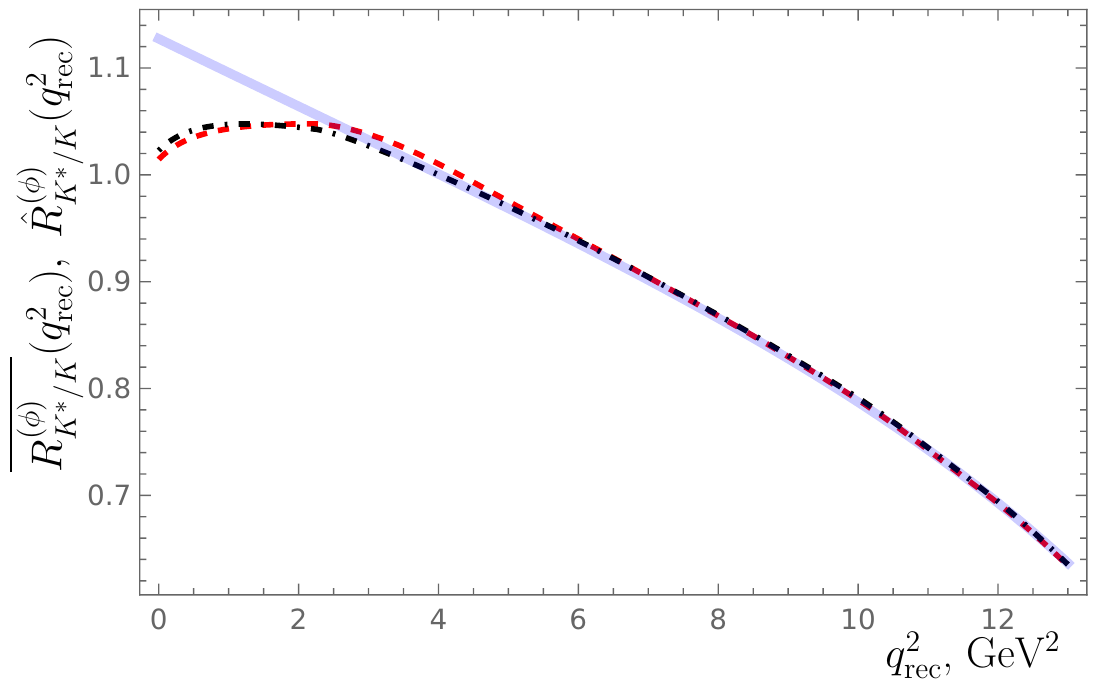}\hspace{.5cm} & \hspace{.5cm}
    \includegraphics[width=6.5cm]{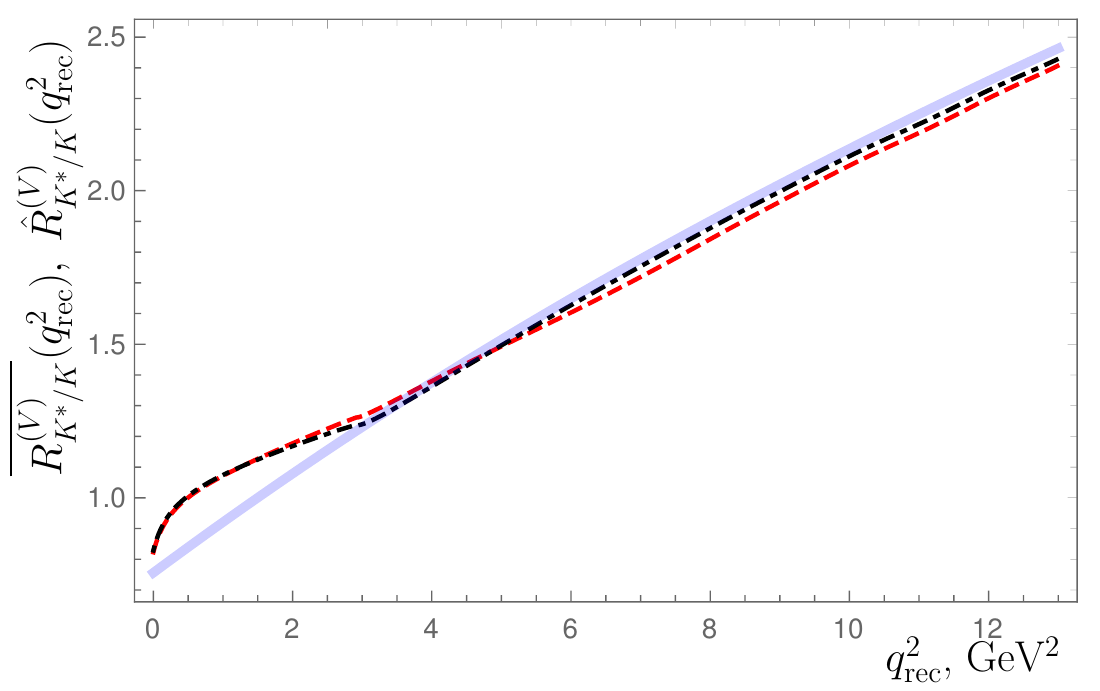}\\
    (a)  &  (b)
\end{tabular}
\caption{
  \label{Fig:6}
 The predicted theoretical ratio $\hat R^{(R)}_{K^*/K}$ [which is independent of DM parameters] (blue solid curve)
  and the predicted ``experimentally measurable'' ratio $\overline{R^{(R)}_{K^*/K}}$ vs.\ $q_\mathrm{rec}^2$
  for S-scenario (a) and V-scenario (b) for two different parameter sets: 
(a) $M_\phi=2.4$~GeV, $\Gamma^0_\phi=2.9$~GeV and $m_\chi=0.42$~GeV (red dashed curve)~and
$M_\phi=20$~GeV, $\Gamma^0_\phi=20$~GeV and $m_\chi=0.42$~GeV (black dotted-dashed curve).
(b) $M_V=3$~GeV, $\Gamma^0_V=4$~GeV and $m_\chi=0.6$~GeV (red dashed curve)~and
$M_V=20$~GeV, $\Gamma^0_V=20$~GeV and $m_\chi=0.6$~GeV (black dotted-dashed curve).}
\end{figure}
\end{center}

\subsection{Measurable vs. theoretical observables}
In Sec.~\ref{Sect:3}
we introduced a number of theoretical observables which may serve as
clear discriminators between DM scenarios. It is time to demonstrate that these observables
remain largely unaffected by the smearing procedures adopted in the data analysis. 

Figure \ref{Fig:6} shows our prediction for the differential distributions of the excess events
in $B\to K^* M_X$ and $B\to K M_X$ decays. Assuming the same detection efficiency for
$B\to K^*M_X$ and $B\to K M_X$ decays,
the ratio $\overline{ R^{(R)}_{K^*/K}}(q_\mathrm{rec}^2)$ of excess events in the
experimentally measured differential distributions in $B\to K^*M_X$ and $B\to K M_X$ decays,
\begin{eqnarray}
\overline{R^{(R)}_{K^*/K}}(q_\mathrm{rec}^2)=\frac{\overline{d\Gamma^{\rm eff}}(B\xrightarrow{R} K^* \chi\bar\chi)/dq_\mathrm{rec}^2}
{\overline{d\Gamma^{\rm eff}}(B\xrightarrow{R} K \chi\bar\chi)/dq_\mathrm{rec}^2},
\end{eqnarray}
(where $\Gamma^{\rm eff}$ indicates that the convolution with the detection efficiencies has been performed,
 see \cite{blm2025}) is practically independent of the model parameters, such as $M_R$, 
$m_\chi$, and $\Gamma^0_R$, as the dependence on these parameters approximately cancels out in the ratio.
(Recall that the dependence on DM parameters cancels exactly in the theoretical ratio $\hat R^{(R)}_{K^*/K}$).

Consequently, we have an approximate equality between the observable ratio
$\overline{R^{(R)}_{K^*/K}}(q_\mathrm{rec}^2)$
and the theoretical ratio $\hat R^{(R)}_{K^*/K}(q_\mathrm{rec}^2)$, which is fulfilled with very high accuracy in a
broad range of momentum transfers (see Fig.~\ref{Fig:6}):
\begin{eqnarray}
\overline{R^{(R)}_{K^*/K}}(q_\mathrm{rec}^2)\simeq \hat R^{(R)}_{K^*/K}(q_\mathrm{rec}^2).
\label{eq:Bstar_B_ratio}
\end{eqnarray}
We therefore conclude that the theoretical observables discussed in Sec.~\ref{Sect:3} indeed provide 
efficient signatures of DM scenarios for data analysis.

\section{\label{Sect:5} Summary and Conclusions} 

We studied conjectures that the observed excess in $B\to K M_X$ is due to the decay $B\xrightarrow{R}K\bar\chi\chi$ into two DM fermions $\bar\chi\chi$, with $R=\phi, V$. 
Our main findings are listed below:

\vspace{.2cm}
\noindent
[i] Within the DM conjecture, ratios of certain observables in $B\to K^*M_X$ and $B\to K M_X$ decays
have qualitatively different behaviors for S- and V-scenarios: 

(a) The ratios of the {\it differential rates} of the excess events in $B\to K^*M_X$ and $B\to K M_X$ decays, 
\begin{eqnarray}
\hat R_{K^*/K}(q^2)=\frac{d\Gamma(B\to K^*\bar\chi\chi)/dq^2}{d\Gamma(B\to K\bar\chi\chi)/dq^2},  
\end{eqnarray}
does not depend on DM parameters, but only on the mediator spin. Measuring this ratio allows one to determine
the mediator spin and to rule out at least one of the DM scenarios. 

(b) The ratio of the {\it integrated rates} of the excess events in $B\to K^*M_X$ and $B\to K M_X$ decays, 
\begin{eqnarray}
R_{K^*/K}=\frac{\Gamma(B\to K^*\bar\chi\chi)}{\Gamma(B\to K\bar\chi\chi)},  
\nonumber
\end{eqnarray}
depends on dark-matter parameters, such as the mediator mass and width, and also has qualitatively different
behavior for S- and V-scenarios. Combining the measured $\Gamma(B\to K M_X)$ and the upper limit
on $\Gamma(B\to K^*M_X)$ provides the allowed regions of the mediator mass within each of the scenarios.
Whereas no constraints on the scalar mediator mass emerge (that is, for any
$M_\phi$ both the measured $\Gamma(B\to K M_X)$ and the upper limit on $\Gamma(B\to K^* M_X)$ are satisfied),
the vector-mediator mass $M_V$ is tightly constrained to the region of small masses, $M_V\lesssim 3$ GeV. 

We emphasize that measuring $\hat R_{K^*/K}(q^2)$ and $R_{K^*/K}$ are excellent discriminators between these two scenarios. 

\vspace{.2cm}
\noindent
[ii] Both S- and V-scenarios allow a good description of the shape of the observed excess events
and extraction of the corresponding DM parameters. A fit to the differential distributions of
the excess events in $B\to K M_X$ yields \cite{blm2025}
\begin{eqnarray}
M_\phi=2.4\pm 0.4\mbox{ GeV}.
\end{eqnarray}
For $M_V$, using the constraint $M_V\lesssim 3$ GeV from the upper limit on the {\it integrated}
rate $\Gamma(B\to K^* M_X)$,
the fitting procedure prefers $M_V$ on the upper boundary of the allowed region, so we just set
\begin{eqnarray}
M_V=3 \mbox{ GeV}
\end{eqnarray}
without providing uncertainties to this number. The values of the other extracted DM parameters are
given in Eqs.~(\ref{DMS}) and (\ref{DMV}).

\vspace{.2cm}
\noindent
[iii] Within {\it any top-philic mediator scenario} one expects a similar excess in 
$B\to (\pi,\rho) M_X$ and $B\to (K,K^*) M_X$ decays 
relative to their SM values: 
\begin{eqnarray}
\frac{\Gamma(B\to \pi M_X)}{\Gamma(B\to \pi\bar\nu\nu)_{\rm SM}}\simeq
\frac{\Gamma(B\to K M_X)}{\Gamma(B\to K\bar\nu\nu)_{\rm SM}} 
\nonumber
\end{eqnarray}
and, similarly, 
\begin{eqnarray}
\frac{\Gamma(B\to \rho M_X)}{\Gamma(B\to \rho\bar\nu\nu)_{\rm SM}}\simeq
\frac{\Gamma(B\to K^* M_X)}{\Gamma(B\to K^*\bar\nu\nu)_{\rm SM}}. 
\nonumber
\end{eqnarray}
(This happens because $\Gamma(B\to \pi\bar\nu\nu)_{\rm SM}$ and 
$\Gamma(B\xrightarrow{R}\pi\bar\chi\chi)$ are proportional to $V_{tb}V^*_{td}$ 
which cancels in the ratios. So, the l.h.s.\ and the r.h.s.\ of the ratios
above are equal to each other up to SU(3)-breaking effects related to $s$- and $d$-quark differences.)

These features are crucial signatures of the top-philic DM scenarios.

\vspace{.2cm}\noindent
{\it Acknowledgments.}
The research was carried out within the framework of the program \emph{Particle Physics and Cosmology}
of the National Center for Physics and Mathematics. The authors thank K\aa{}re Fridell for helpful
comments on the study~\cite{Fridell:2023ssf}.



\begin{thebibliography}{50}%

\bibitem{damir2023epjc}
D.~Becirevic, G.~Piazza, and O.~Sumensari,
{\it Revisiting $B\to K^{(*)}\nu\bar\nu$ decays in the Standard Model and beyond},
Eur.~Phys.~J.~C {\bf 83}, 252 (2023).
\bibitem{hpqcd2023}
W.~G.~Parrott, C.~Bouchard, and C.~T.~H.~Davies [HPQCD collaboration],
{\it Standard Model predictions for $B\to Kl^+l^-$, $B\to K l_1^-l_2^+$, and $B\to K \nu\bar\nu$
using form factors from $N_f = 2 + 1 + 1$ lattice QCD},
Phys.~Rev.~D {\bf 107}, 014511 (2023); {\bf 107}, 119903(E) (2023).
\bibitem{Allwicher:2023xba}
 L. Allwicher, D. Becirevic, G. Piazza, S. Rosauro-Alcaraz, and O. Sumensari,
 {\it Understanding the first measurement of $B(B\to K\nu\nu)$},
 Phys. Lett. B {\bf 848}, 138411 (2024), arXiv:2309.02246 [hep-ph].
\bibitem{belle2017}
J. Grygier et al. (Belle Collaboration),
{\it Search for $B\to h\nu\bar\nu$ decays with semileptonic tagging at Belle}, 
Phys.~Rev.~D {\bf 96}, 091101 (2017). 
\bibitem{Belle-II:2023esi}
 I. Adachi et al. (Belle-II Collaboration),
 {\it Evidence for $B^+\to K^+\nu\nu$ decays}, Phys. Rev. D {\bf 109}, 112006 (2024),
 arXiv:2311.14647 [hep-ex].
\bibitem{Athron:2023hmz}%
 P. Athron, R. Martinez, and C. Sierra,
 {\it B meson anomalies and large $B\to K\nu\nu$ in non-universal $U(1)'$ models},
 JHEP {\bf 02}, 121 (2024), arXiv:2308.13426 [hep-ph].
\bibitem{Bause:2023mfe}%
 R. Bause, H. Gisbert, and G. Hiller,
 {\it Implications of an enhanced $B\to K\nu\nu$ branching ratio},
 Phys. Rev. D {\bf 109}, 015006 (2024), arXiv:2309.00075 [hep-ph].
\bibitem{Felkl:2023ayn}%
 T. Felkl, A. Giri, R. Mohanta, and M. A. Schmidt,
 {\it When energy goes missing: new physics in $b\to s\nu\nu$ with sterile neutrinos},
  Eur. Phys. J. C {\bf 83}, 1135 (2023), arXiv:2309.02940 [hep-ph].
\bibitem{Abdughani:2023dlr}%
 M. Abdughani and Y. Reyimuaji,
 {\it Constraining light dark matter and mediator with $B\to K\nu\nu$ data},
 Phys. Rev. D {\bf 110}, 055013 (2024), arXiv:2309.03706 [hep-ph].

\bibitem{Dreiner:2023cms}%
 H. K. Dreiner, J. Y. Guenther, and Z. S. Wang,
 {\it The Decay $B\to K\nu\nu$ at Belle II and a Massless Bino in R-parity-violating Supersymmetry},
   Eur. Phys. J. C {\bf 85}, 66 (2025), arXiv:2309.03727 [hep-ph]

\bibitem{He:2023bnk}%
 X.-G. He, X.-D. Ma, and G. Valencia,
 {\it Revisiting models that enhance $B\to K\nu\nu$ in light of the new Belle II measurement},
Phys. Rev. D {\bf 109}, 075019 (2024), arXiv:2309.12741 [hep-ph].

\bibitem{Berezhnoy:2023rxx}
A.~Berezhnoy and D.~Melikhov,
{\it $B\to K^* M_X$ vs $B\to K M_X$ as a probe of a scalar-mediator dark matter scenario}, 
EPL \textbf{145}, 14001 (2024), arXiv:2309.17191 [hep-ph].

\bibitem{datta}%
A.~Datta, D.~Marfatia, and L.~Mukherjee,   
{\it $B\to K\bar nu\nu$ MiniBooNE and muon $g-2$ anomalies from a dark sector},
Phys. Rev. D {\bf 109}, L031701 (2024), arXiv:2310.15136 [hep-ph].

\bibitem{Calibbi:2025rpx}%
 L. Calibbi, T. Li, L. Mukherjee, and M. A. Schmidt,
 {\it Is Dark Matter the origin of the $B\to K\nu\nu$ excess at Belle II?}, arXiv:2502.04900 [hep-ph].

\bibitem{Altmannshofer:2025eor}%
 W. Altmannshofer, S. A. Gadam, and K. Toner,
 {\it New Strategies for New Physics Search with $\Lambda_b\to \Lambda \nu\nu$ Decays},
 arXiv:2501.10652 [hep-ph].

\bibitem{Zhang:2024hkn}%
 C.-Q. Zhang, J. Sun, Z.-P. Xing, and R.-L. Zhu,
 {\it Probing $B\to K^{(*)}$ semi-leptonic FCNC decay with new physics effects under PQCD approach},
 arXiv:2501.00512 [hep-ph].

\bibitem{Bhattacharya:2024clv}%
 B. Bhattacharya, A. Datta, G. Faisel, S. Khalil, and S. Roy,
 {\it Flavor Violations in B-Mesons within Non-Minimal SU(5)},
 arXiv:2412.16115 [hep-ph].
\bibitem{Altmannshofer:2024kxb}%
 W. Altmannshofer and S. Roy,
 {\it A joint explanation of the $B\to\pi K$ puzzle and the $B\to K\nu\nu$ excess},
 arXiv:2411.06592 [hep-ph].
\bibitem{Davoudiasl:2024cee}%
 H. Davoudiasl and M. Schnubel,
 {\it Bringing the Peccei-Quinn mechanism down to Earth},
 Phys. Rev. D {\bf 110}, 075014 (2024),
 arXiv:2406.19455 [hep-ph].
\bibitem{Kim:2024tsm}%
 C. S. Kim, D. Sahoo, and K. N. Vishnudath,
 {\it Searching for signatures of new physics in $B\to K\nu\nu$ to distinguish between Dirac and Majorana neutrinos},
 Eur. Phys. J. C {\bf 84}, 882 (2024), arXiv:2405.17341 [hep-ph
\bibitem{Marzocca:2024hua}%
 D. Marzocca, M. Nardecchia, A. Stanzione, and C. Toni,
 {\it Implications of $B\to K\nu\nu$ under rank-one flavor violation hypothesis},
 Eur. Phys. J. C {\bf 84}, 1217 (2024), arXiv:2404.06533 [hep-ph].
\bibitem{Bolton:2024egx}%
 P. D. Bolton, S. Fajfer, J. F. Kamenik, and M. Novoa-Brunet,
 {\it Signatures of light new particles in $B\to K^{(*)}E_{\rm miss}$},
  Phys. Rev. D {\bf 110}, 055001 (2024), arXiv:2403.13887 [hep-ph].
\bibitem{He:2024iju}%
 X.-G. He, X.-D. Ma, M. A. Schmidt, G. Valencia, and R. R. Volkas,
 {\it Scalar dark matter explanation of the excess in the Belle II $B\to K +$ invisible measurement},
 JHEP {\bf 07}, 168 (2024), arXiv:2403.12485 [hep-ph].
\bibitem{Hou:2024vyw}%
 B.-F. Hou, X.-Q. Li, M. Shen, Y.-D. Yang, and X.-B. Yuan,
 {\it Deciphering the Belle II data on $B\to K\nu\nu$ decay in the (dark)
SMEFT with minimal flavour violation}, JHEP {\bf 06}, 172 (2024), arXiv:2402.19208 [hep-ph].
\bibitem{Gabrielli:2024wys}%
 E. Gabrielli, L. Marzola, K. Muursepp, and M. Raidal,
 {\it Explaining the $B\to K\nu\nu$ excess via a massless dark photon},
Eur. Phys. J. C {\bf 84}, 460 (2024), arXiv:2402.05901 [hep-ph].
\bibitem{Loparco:2024olo}%
 F. Loparco, {\it A New Look at $b\to s$ Observables in 331 Models},
 Particles {\bf 7}, 161 (2024), arXiv:2401.11999
\bibitem{Ho:2024cwk}%
 S.-Y. Ho, J. Kim, and P. Ko,
 {\it Recent $B\to K\nu\nu$ Excess and Muon $g-2$ Illuminating Light Dark Sector with Higgs Portal},
 arXiv:2401.10112 [hep-ph].
\bibitem{Fridell:2023ssf}%
 K. Fridell, M. Ghosh, T. Okui, and K. Tobioka,
 {\it Decoding the $B\to K\nu\nu$ excess at Belle II: Kinematics, operators, and masses},
 Phys. Rev. D {\bf 109}, 115006 (2024), arXiv:2312.12507 [hep-ph].
\bibitem{McKeen:2023uzo}%
 D. McKeen, J. N. Ng, and D. Tuckler, {\it Higgs portal interpretation of the Belle II $B\to K\nu\nu$ measurement},
 Phys. Rev. D {\bf 109}, 075006 (2024), arXiv:2312.00982 [hep-ph].
\bibitem{Altmannshofer:2023hkn}%
 W. Altmannshofer, A. Crivellin, H. Haigh, G. Inguglia, and J. Martin Camalich,
 {\it Light new physics in $B\to K^{(*)}\nu\nu$?},
 Phys. Rev. D {\bf 109}, 075008 (2024), arXiv:2311.14629 [hep-ph].
\bibitem{Wang:2023trd}%
 Z. S.Wang, H. K. Dreiner, and J. Y. Guenther,
 {\it The decay $B\to K\nu\nu$ at Belle II and a massless bino in R-parity-violating
 supersymmetry}, Eur. Phys. J. C {\bf 85}, 66 (2025), arXiv:2309.03727 [hep-ph].
\bibitem{He:2025jfc}
X.~G.~He, X.~D.~Ma, J.~Tandean and G.~Valencia,
{\it $B\to K\sf{+} \mathrm{invisible}$, dark matter, and $CP$ violation in hyperon decays}, arXiv:2502.09603 [hep-ph].
\bibitem{fajfer}
P.~D.~Bolton, S.~Fajfer, J.~F.~Kamenik, and M.~Novoa-Brunet, 
{\it Impact of new invisible particles on $B\to K(*)Emiss$ observables}, 
arXiv:2503.19025 [hep-ph]

\bibitem{blm2025}%
  A. Berezhnoy, W. Lucha, and D. Melikhov,
  {\it Analysis of $q^2$-distribution for $B\to K M_X$ and $B\to K^* M_X$
    in a scalar-mediator dark matter scenario},
  Phys. Rev. D {\bf 111}, 075035 (2025), arXiv:2502.14313 [hep-ph].
\bibitem{Kolay:2025jip}
L.~Kolay and S.~Nandi,
{\it Flavour and Electroweak Precision Constraints on a Simplified Dark Matter Model with a Light Spin-0 Mediator}, 
[arXiv:2503.15609 [hep-ph]].
\bibitem{Aliev:2025hyp}
T.~M.~Aliev, A.~Elpe, I.~Turan and L.~Selbuz,
{\it Explaining Belle Data on $B\to K^{(*)}\nu\bar{\nu}$ Decays via Dark Photon Resonances}, 
arXiv:2503.22347 [hep-ph].
\bibitem{Ding:2025eqq}
K.~Ding, Y.~Li, X.~Liu, Y.~Liu, C.~T.~Lu and B.~Zhu,
{\it Resonant ALP-Portal Dark Matter Annihilation as a Solution to the $B^{\pm} \to K^{\pm} \nu \bar{\nu}$ Excess,} 
arXiv:2504.00383 [hep-ph].

\bibitem{He:2025zfy}
X.~G.~He, X.~D.~Ma, J.~Tandean and G.~Valencia,
{\it Light dark-matter window in $K^+\to\pi^+$$+$$\not{\!\!E}$}, 
arXiv:2505.02031 [hep-ph].

\bibitem{DiLuzio:2025whh}
L.~Di Luzio, M.~Nardecchia and C.~Toni,
{\it Gauged $\tau$-lepton chiral currents and $B \to K^{(*)} E_{\rm miss}$}, 
arXiv:2505.11499 [hep-ph].



\bibitem{Batell:2009jf}%
 B. Batell, M. Pospelov, and A. Ritz,
 {\it Multi-lepton Signatures of a Hidden Sector in Rare B Decays}, Phys. Rev. D {\bf 83},
054005 (2011), arXiv:0911.4938 [hep-ph].
\bibitem{Schmidt-Hoberg:2013hba}%
 K. Schmidt-Hoberg, F. Staub, and M. W. Winkler,
 {\it Constraints on light mediators: confronting dark matter searches with
B physics}, Phys. Lett. B {\bf 727}, 506 (2013), arXiv:1310.6752 [hep-ph].
\bibitem{Langacker:2008yv}
P.~Langacker,
{\it The Physics of Heavy $Z^\prime$ Gauge Bosons}, 
Rev.~Mod.~Phys. \textbf{81}, 1199 (2009)
arXiv:0801.1345 [hep-ph].
\bibitem{Cox:2015afa}
P.~Cox, A.~D.~Medina, T.~S.~Ray and A.~Spray,
{\it Novel collider and dark matter phenomenology of a top-philic $Z'$}, 
JHEP \textbf{06}, 110 (2016)
arXiv:1512.00471 [hep-ph].
\bibitem{Hu:2024xes}
Y.~Hu, Y.~Liu and Y.~Liu,
{\it A study on vector mediator top-philic dark matter}, 
Commun.~Theor.~Phys. \textbf{76}, 085201 (2024)
\bibitem{Inami:1980fz}
T.~Inami and C.~S.~Lim,
{\it Effects of Superheavy Quarks and Leptons in Low-Energy Weak Processes
$K_L\to \mu\bar\mu$, $K^+\to \pi^+\nu\bar\nu$ and $K^0\leftrightarrow \bar K^0$}, 
Prog.~Theor.~Phys.~\textbf{65}, 297 (1981)
[erratum: Prog.~Theor.~Phys.~\textbf{65}, 1772 (1981)]
\bibitem{Melikhov:1997wp}
D.~Melikhov, N.~Nikitin and S.~Simula,
{\it Rare exclusive semileptonic $b\to s$ transitions in the standard model}, 
Phys.~Rev.~D~\textbf{57}, 6814 (1998), 
arXiv:hep-ph/9711362 [hep-ph].
\bibitem{gs}G. J. Gounaris and J. J. Sakurai, 
{\it Finite width corrections to the vector meson dominance prediction for $\rho\to e^+e^-$}, 
Phys.~Rev.~Lett.~{\bf 21}, 244 (1968). 
\bibitem{nachtmann}
D. Melikhov, O. Nachtmann, V. Nikonov, and T. Paulus,
{\it Masses and couplings of vector mesons from the pion electromagnetic, weak, and pi gamma transition form-factors}, Eur.~Phys.~J.~C~{\bf 34}, 345 (2004).
\bibitem{wsb}
M.~Wirbel, B.~Stech, and M.~Bauer, {\it Exclusive Semileptonic Decays of Heavy Mesons},
Z.~Phys.~C {\bf 29}, 637 (1985).

\bibitem{Bailey:2015dka}
J.~A.~Bailey, A.~Bazavov, C.~Bernard, C.~M.~Bouchard, C.~DeTar, D.~Du, A.~X.~El-Khadra, J.~Foley, E.~D.~Freeland and E.~G{\'a}miz, \textit{et al.}
{\it $B\to Kl^+l^-$ Decay Form Factors from Three-Flavor Lattice QCD}, 
Phys.~Rev.~D \textbf{93}, 025026 (2016)
[arXiv:1509.06235 [hep-lat]].
\bibitem{ms2000}
D.~Melikhov and B.Stech,
{\it Weak form-factors for heavy meson decays: an update},
Phys.~Rev.~D {\bf 62}, 014006 (2000).

\bibitem{Bharucha:2015bzk}
A.~Bharucha, D.~M.~Straub and R.~Zwicky,
{\it $B\to V\ell^+\ell^-$ in the Standard Model from light-cone sum rules}, 
JHEP \textbf{08}, 098 (2016)
[arXiv:1503.05534 [hep-ph]].

\bibitem{Khodjamirian:2010vf}
A.~Khodjamirian, T.~Mannel, A.~A.~Pivovarov and Y.~M.~Wang,
{\it Charm-loop effect in $B \to K^{(*)} \ell^{+} \ell^{-}$ and $B\to K^*\gamma$}, 
JHEP \textbf{09}, 089 (2010)
[arXiv:1006.4945 [hep-ph]].





\end{thebibliography}
\end{document}